%% file: manuscript.tex
\documentclass[review]{elsarticle}

\usepackage{amssymb}
\usepackage{amsmath}
\usepackage{mathtools}
\usepackage{multirow}
\usepackage{subfig}
\usepackage{bm}
\usepackage{color, soul}
\usepackage[font={footnotesize}]{caption}
\usepackage{array}
\usepackage{paralist}
\usepackage[dvipsnames]{xcolor}

\newcommand{\newlinecell}[2][c]{%
  \begin{tabular}[#1]{@{}c@{}}#2\end{tabular}}

\newcolumntype{C}[1]{>{\centering\let\newline\\\arraybackslash\hspace{0pt}}m{#1}}
\usepackage{lineno,hyperref}
\modulolinenumbers[5]

\journal{Pattern Recognition}

\begin{document}

\begin{frontmatter}

\title{Learning sound representations using \\trainable COPE feature extractors}

\author[rug]{Nicola Strisciuglio}
\cortext[mycorrespondingauthor]{Corresponding author: n.strisciuglio@rug.nl}
\author[unisa]{Mario Vento}
\author[rug]{Nicolai Petkov}

\address[rug]{Bernoulli Institute for Mathematics, Computer Science and Artificial Intelligence - University of Groningen, The Netherlands.}
\address[unisa]{Department of Information and Electrical Engineering and Applied Mathematics (DIEM) - University of Salerno, Italy.}

\begin{abstract}
Sound analysis research has mainly been focused on speech and music processing. The deployed methodologies are not suitable for analysis of sounds with varying background noise, in many cases with very low signal-to-noise ratio (SNR).

In this paper, we present a method for the detection of patterns of interest in audio signals. We propose novel trainable feature extractors, which we call COPE (Combination of Peaks of Energy). The structure of a COPE feature extractor is determined using a single prototype sound pattern in an automatic configuration process, which is a type of representation learning. We construct a set of COPE feature extractors, configured on a number of training patterns. Then we take their responses to build feature vectors that we use in combination with a classifier to detect and classify patterns of interest in audio signals. 

We carried out experiments on four public data sets: MIVIA audio events, MIVIA road events, ESC-10 and TU Dortmund data sets. The results that we achieved (recognition rate equal to ${91.71 \%}$ on the MIVIA audio events, ${94 \%}$ on the MIVIA road events, ${81.25 \%}$ on the ESC-10 and ${94.27 \%}$ on the TU Dortmund) demonstrate the effectiveness of the proposed method and are higher than the ones obtained by other existing approaches. The COPE feature extractors have high robustness to variations of SNR. Real-time performance is achieved even when the value of a large number of features is computed.
\end{abstract}

\begin{keyword}
audio analysis, event detection, peaks of energy, representation learning, trainable feature extractors 
\end{keyword}

\end{frontmatter}


\graphicspath{ {./figures/}
				{./figures/authors/}
				{./figures/method/}
				{./figures/method/gammatone/}
				{./figures/discussion/}
				{./figures/comparison/}
				{./figures/comparison/ESC10/}
				{./figures/results/}}


\section{Introduction}
\label{sec:introduction}
Methods and systems for the automatic analysis of people and vehicle behavior, scene understanding, familiar place recognition and human-machine interaction are traditionally based on computer vision techniques. In robotics or public security, for instance, there has been a great effort to equip machines with capabilities for autonomous visual understanding. However, video analysis has some weak points, such as sensitivity to light changes and occlusions, or limitation  to the field of view of the camera. Sound is complementary to visual information and can be used to improve the capabilities of machines to deal with the surrounding environment. Furthermore, there are  cases in which video analysis cannot be used due to privacy issues (e.g. in public toilets). 

In this paper we focus on automatic learning of  representations of sounds that are suitable for pattern recognition, in the context of environmental sound analysis for detection and classification of audio events. 
Recently, the interest in automatic analysis of environmental sounds increased because of  various applications in intelligent surveillance and security~\cite{CroccoSurvey}, assistance of eldery people~\cite{Vacher2010}, monitoring of smart rooms~\cite{Wang2008}, home and social robotics~\cite{Maxime2014}, etc.

A large part of sound analysis research in the past years focused on  speech recognition~\cite{Besacier201485}, speaker identification~\cite{Roy2012} and music classification~\cite{Fu2011survey}. 
Features and classifiers for voice analysis are established and widely used in practical systems: spectral or cepstral features in combination with classifiers based on Hidden Markov Models or Gaussian Mixture Models.  
However, state of the art methods for speech and music analysis do not give good results when applied to environmental sounds, which have highly non-stationary characteristics~\cite{Cowling2003}. 
Most speech recognition methods assume that  speech is based on a phonetic structure, which allows to analyze complex words or phrases by splitting them in a series of simple phonemes. In the case of environmental sound there is no underlying phoneme-like structure.
Moreover, human voice has very specific frequency characteristics that are not present in other kinds of sound. For example, interesting events for surveillance applications, such as gun shots or glass breaking  usually have high-frequency components that are not present in speech. 
For speech recognition and speaker identification the sound source is typically very close to the microphone. It implies that  background noise has lower energy  than foreground sounds and does not impair considerably the performance of the recognition system. 
Environmental sound sources can be, instead, at any distance from the microphone. Hence, the background noise can have relatively high energy, so determining very low or even negative signal-to-noise ratio (SNR).

Existing methods for detection of audio events, for which we provide an extensive overview in Section~\ref{sec:literature}, are based on the extraction of hand-crafted features from the audio signal. The features extracted from (a part of) the audio signal are submitted to a classification system. The employed features describe stationary and non-stationary properties of the signals~\cite{Chachada2013}.
This approach to pattern recognition requires a feature engineering step that aims at choosing or designing a  set of features that describe important characteristics of the sound for the problem at hand. Widely used features are mainly borrowed from the field of speech recognition: responses of log-frequency filters, Mel-frequency cepstral coefficients, wavelet transform coefficients among others. The choice of effective features or combination of them is a critical step to build an effective system and requires considerable domain knowledge.

More recent approaches do not rely on hand-crafted features but rather involve automatic learning of data representations from training samples by using deep learning and convolutional neural networks (CNN)~\cite{lecun2015deep}.
CNNs were originally proposed for visual data analysis, but have also been successfully applied to speech~\cite{Huang2015}, music processing~\cite{Oord2013} and sound scene classification~\cite{Phan2017}. While they achieve very good performance, they require very large amount of labeled training data which is not always available. 

In this work, we propose trainable feature extractors for sound analysis which we call COPE (Combination of Peaks of Energy). They are trainable as their structure is not fixed in advance but it is rather learned in an automatic configuration procedure using a single prototype pattern. 
This automatic configuration of feature extractors is a type of  \emph{representation learning}. It allows to automatically construct a suitable data representation to be used together with a classifier and does not require considerable domain expertise.
We configure a number of COPE feature extractors on training sounds and use their responses to build a feature vector, which we then employ as input to a classifier. With respect to~\cite{CopePreliminary2015}, in which we reported preliminary results obtained using COPE feature extractors on sound events with the same SNR, in this work we provide: \begin{inparaenum}[\itshape a\upshape)]
\item a detailed formulation of the configuration and application steps of COPE features,
\item a thourough validation of the performance of a classification system based on COPE features when tested with sounds with different values of SNR,
\item an extension of the MIVIA audio events data set the includes null or negative SNR sound events and 
\item a wide comparison of the proposed method with other existing approaches on four benchmark data sets.
\end{inparaenum}
Furthermore, we discuss the importance of robustness to variations of the background noise and SNR of the events of interest, for applications of sound event detection in Section~\ref{sec:snr}. We provide a detailed analysis of the contribution of the COPE features to the improvement of sound event detection and classification performance with respect to existing approaches.

The design of COPE feature extractors was inspired by certain properties of the inner auditory system, which converts the sound pressure waves that reach our ears into neural stimuli on the auditory nerve. In the~\ref{sec:motivation} we provide some details about the biological mechanisms that inspired the design of the COPE feature extractors.

We validate the effectiveness of the proposed COPE feature extractors by carrying out experiments on the following public benchmark data sets: MIVIA audio events~\cite{BoawPRL2015},  MIVIA road events~\cite{ITS2015}, ESC-10~\cite{Piczak15}, TU-Dortmund~\cite{Plinge2014}. 

The main contributions of this work are: 
\begin{inparaenum}[\itshape a\upshape)]
\item novel COPE trainable feature extractors for representation learning of sounds that are automatically configured on training examples, 
\item a method for audio event detection that uses the proposed features, 
\item the release of an extended version of the MIVIA audio events data set with sounds at null and negative SNR.
\end{inparaenum}

The rest of the paper is organized as follows. In Section~\ref{sec:literature} we review related works, while in Section~\ref{sec:method} we present the COPE feature extractors and the architecture of the proposed method. We describe the data sets used for the experiments in Section~\ref{sec:dataset}. We report the results that we achieved, a comparison with existing methods and an analysis of the sensitivity of the performance of the proposed method with respect to the parameters of the COPE feature extractors in Section~\ref{sec:experiments}. We provide a discussion in Section~\ref{sec:discussion} and, finally, draw conclusions in Section~\ref{sec:conclusion}.

\section{Related works}
\label{sec:literature}
 
Representation learning has recently received great attention by researchers in pattern recognition with the aim of constructing reliable features by direct learning from training data. Methods based on deep learning and CNNs were proposed to learn features for several applications: age and gender estimation from facial images~\cite{LIU201782}, action recognition~\cite{Deep2014}, person re-identification~\cite{DING20152993}, hand-written signature verification~\cite{HAFEMANN2017163}, and also sound analysis~\cite{Aytar2016}. Other approaches for feature learning focused on sparse dictionary learning~\cite{Rubinstein09,CHEN201751}, learning vector quantization~\cite{BUNTE20111892}, and on extensions of the bag of features approach based on neural networks~\cite{PASSALIS2017277} or higher-order pooling~\cite{Koniusz17}.

In the context of audio analysis research, it is common to organize existing works on sound event detection according to the feature sets and classification architectures that they employ. Early methods approached the problems of sound event detection and classification by dividing the audio signal into small, partially overlapped frames and computing a feature vector for each frame. The used features  ranged from relatively simple (e.g. frame energy, zero-crossing rate, sub-band energy rate) to more complicated ones (e-g. Mel-frequency Cepstral Coefficients~\cite{Guo2003}, log-frequency filter banks~\cite{Nadeu2001}, perceptual linear prediction coefficients~\cite{Portelo2009}, etc.). 
The frame-level feature vectors were then used together with a classifier to perform a decision. Gaussian Mixture Model (GMM) based classifiers were largely employed to classify the frames as part of sounds of interest or background~\cite{clavel05,Atrey2006}.
To limit the influence of background sounds on the classification performance, One-Class Support Vector Machines were proposed~\cite{Rabaoui2008}. 

Spectro-temporal features based on spectrogram or other time-frequency representations were also developed~\cite{Chu2009, Dennis13}. Inspired by the way the inner auditory system of humans responds to the frequency of the sounds, an auditory image model (AIM) was proposed~\cite{Patterson1992}. The AIM was used as basis for improved models which are called stabilized auditory images (SAI)~\cite{Lyon2011}.
In~\cite{gammatoneAVSS}, the event detection was formulated as an object detection problem in a spectrogram-like representation of the sound, and approached by using a cascade of AdaBoost classifiers. The design of hand-crafted features poses some limitations to the construction of systems that are robust to varying conditions of the events of interest and requires considerable domain knowledge.

In order to construct more reliable systems, efforts towards automatic learning of features from training data by means of machine learning techniques were made. Various approaches based on  \emph{bag of features}  were proposed for sound event representation and classification~\cite{Aucouturier07,Pancoast12}. A code-book of basic audio features (also called \emph{audio words}) is directly learned from training samples as result of a quantization of the feature space by means of various clustering algorithms (e.g. \emph{k}-Means or fuzzy \emph{k}-Means). A comparison of hard and soft quantization of audio words was performed in~\cite{BoawPRL2015}. Other approaches for the construction of a code-book of basic audio words were also based on non-negative matrix factorization~\cite{Giannoulis13} or sparse coding~\cite{Lu2014}.
In the bag of features representation, the information about the temporal arrangement of the audio words is lost.  This was taken into account in~\cite{Grzeszick15} and~\cite{chin2012}, where a feature augmentation and a classifier based on Genetic Motif Discovery were proposed, respectively. The sequence of audio words were also employed in~\cite{Kumar12} and~\cite{PhanM15}. The temporal information was described by a pyramidal approach to bag of features in~\cite{Plinge2014,Grzeszick2016}.
A method for sound representation learning based on Convolutional Neural Networks (CNN) was proposed in~\cite{Piczak15a}. Learning features from training samples does not require an engineering effort and allows for the adaptation of the recognition systems to various problems. However, the effectiveness and generalization capabilities of learned features depend on the amount of available training data.



Evaluation of algorithms for audio event detection on public benchmark data sets is a valuable tool for objective comparison of performance. 
The great attention that was dedicated to music and speech analysis determined the publication of several data sets used in scientific challenges for benchmarking of algorithms. The MIREX challenge series evaluated systems for music information retrieval (MIR)~\cite{Downie2010}. The CHiME challenge focused on speech analysis in noisy environments~\cite{CHiME13}. 
The ``Acoustic event detection and classification'' task of the CLEAR challenges (2006 and 2007) focused on the detection of sound events related to seminars, such as speech, chair moving, door opening and applause~\cite{Stiefelhagen2007}. Recently, the DCASE challenge~\cite{Stowell2015} stimulated the interest of researchers on audio processing for the analysis of environmental sounds. The attention was driven towards audio event detection and classification and scene classification. 


\section{Method}
\label{sec:method}

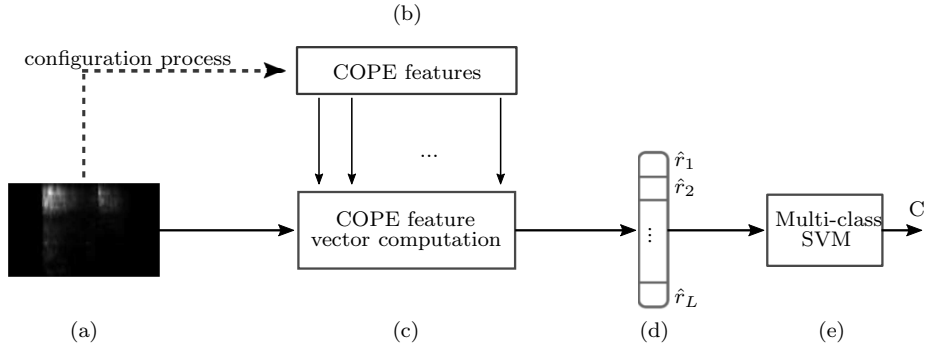
\begin{figure}[!t]
	\centering
	\footnotesize
   \input{figures/method/schema-paper.eps_tex}
   \caption{Architecture of the proposed method. The (a) Gammatonegram of the training audio samples is computed in the training phase (dashed arrow), and used to configure a (b) set of COPE feature extractors. The learned features are used in the application phase to (c) process the input sound and (d) construct feature vectors with their responses. A (e) multi-class SVM classifier is, finally, employed to detect events of interest.}
   \label{fig:schema-cope}
\end{figure}

In Figure~\ref{fig:schema-cope}, we show an overview of the architecture of the proposed method. The algorithm is divided in two phases: configuration and application.

In the configuration phase (dashed line), the Gammatonegrams (see details in Section~\ref{sec:gammatonegram}) of prototype training sounds are used to configure a set of COPE feature extractors (see Section~\ref{sec:method-conf}). Successively, the response of the set of COPE feature extractors, computed on the sounds in the training set, are employed to construct COPE feature vectors (Figure~\ref{fig:schema-cope}b-d). A multi-class SVM classifier is finally trained using the COPE feature vectors (Figure~\ref{fig:schema-cope}e) to distinguish between the classes of interest for the application at end. 

In the application phase, the previously configured set of COPE feature extractors is applied to extract feature vectors from input unknown sounds and the multi-class SVM classifier is used to detect and classify sound events of interest.  
The implementation of the COPE feature extractors and the proposed classification architecture is publicly available\footnote{The code is available at http://gitlab.com/nicstrisc/COPE}.





\subsection{Gammatonegram}
\label{sec:gammatonegram}
The traditional and most used time-frequency representation of sounds is the spectrogram, in which the energy distribution over frequencies is computed by dividing the frequency axis into sub-bands with equal bandwidth. 
In the human auditory system, the resolution in the perception of differences in frequency changes according to the base frequency of the sound. At low frequency the band-pass filters have a narrower bandwidth than the ones at high frequency. This implies higher time resolution of filters at high frequency that are able to better catch high variations of the signal. In this work we employ a bank of Gammatone band-pass filters, whose bandwidth increases with increasing central frequency. 
The functional form of Gammatone is biologically-inspired and models the response of the cochlea membrane in the inner ear of the human auditory system~\cite{patterson1986auditory}. 

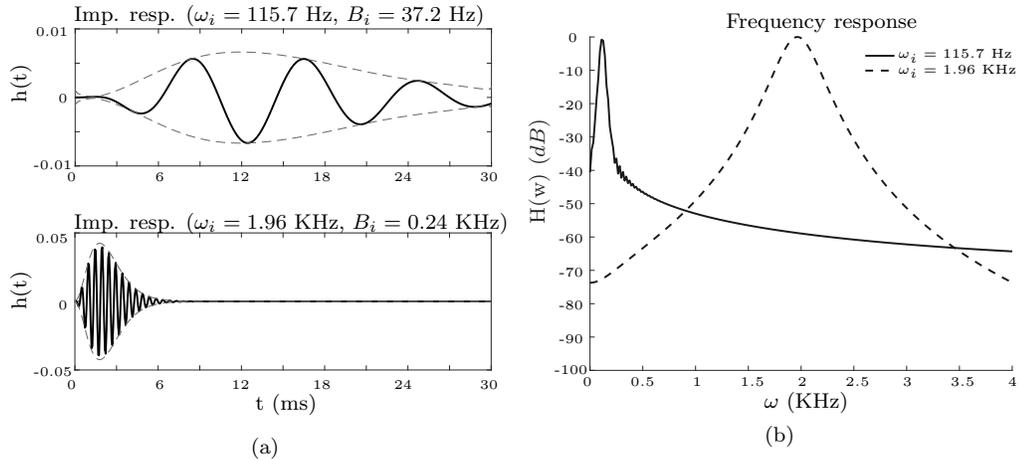
\begin{figure}[!t]
	\centering
\begin{footnotesize}
   \subfloat[]{ 	\label{fig:impulse}
   	\begin{tabular}[b]{l}
   		\input{figures/method/gammatone/impulse-low-paper.eps_tex}
\\
   \input{figures/method/gammatone/impulse-high-paper.eps_tex}
   \end{tabular}
   }
~
\subfloat[]{\label{fig:freq-resp}
\input{figures/method/gammatone/frequency-resp-paper.eps_tex}}
\end{footnotesize}
   \caption{(a) Impulse responses of two Gammatone filters (with central frequencies $\omega_1=115.1$ Hz and $\omega_2=1.96$ KHz). The dashed lines represent the envelope of the sinusoidal tone. The (b) frequency responses of the filters in (a): the filter with higher central frequency (dashed line) has larger bandwidth ($B_2=240$ Hz while $B_1=37.2$ Hz).}
   \label{fig:gt-examples}
\end{figure}

The impulse response of a Gammatone filter is the product of a statistical distribution called \emph{Gamma} and a sinusoidal carrier tone. It is formally defined as:

\begin{equation}
h_i(t) = 
\begin{cases}
 	at^{n-1} e^{-2 \pi B_i t} cos(2 \pi \omega_i t + \phi), & t \geq 0\\ 
    0,& else
\end{cases},
\end{equation}
where $\omega_i$ is the central frequency of the filter, and $\phi$
is its phase. The constant $a$ controls the
gain and $n$ is the order of the filter. The parameter $B_i$ is a decay factor and determines the bandwidth of the band-pass filter.
The relation between the central frequency of a Gammatone filter and its bandwidth is given by the Equivalent Rectangular Bandwidth (ERB):
\begin{equation}
B_i = \left[ \left( \frac{\omega_i}{Q_{ear}} \right) ^ p + \left( B_{min} \right) ^p \right] ^ {1/p}
\end{equation}
\noindent where $Q_{ear}$ is the asymptotic filter quality at high frequencies and $B_{min}$ 
is the minimum bandwidth at low frequencies, while $p$ is usually equal to 1 or 2. In~\cite{Glasberg1990}, the parameters $Q_{ear} = 9.26779$, $B_{min} = 24.7$ and $p=1$ where determined by measurements from notched-noise data. 
In Figure~\ref{fig:impulse}, we show the impulse response of two Gammatone filters with low ($\omega_1=115.1$ Hz) and higher ($\omega_2=1.96$ KHz) central frequencies. The filter with higher central frequency has larger bandwidth, as it can be seen from their frequency response in Figure~\ref{fig:freq-resp}.

We filter the input signal $x(t)$ with a bank of $\Gamma$ Gammatone filters $\underline{\mathbf{h}}(t) = \left[ h_1(t), h_2(t), \dots, h_\Gamma (t) \right] ^ T$. The response of the $i$-th filter to an input signal $x(t)$ is the convolution of the input signal with the impulse response $h_i(t)$:
\begin{equation}
	\tilde{x}_i(t) = x(t) \ast h_i(t).
\end{equation}

We divide the input audio signal in frames of $F$ samples and process every frame by a bank of Gammatone filters in order to capture the short-time properties of the energy distribution of the sound. Two consecutive frames have $F / 2$ samples in common, which means that they overlap for $50\%$ of their length. This ensures continuity of analysis and that border effects are avoided. Given an input signal with $N$ samples, the number of concerned frames is  $\Theta = \lfloor 2(N-F)/F \rfloor + 1$.
We finally construct the Gammatonegram of a sound as a matrix $\mathcal{X} \in \mathcal{R}^{\Gamma \times \Theta}$, whose $j$-th column corresponds to  $\left[ \tilde{x}_i(jF/2), \tilde{x}_i(jF/2 + 1), \dots, \tilde{x}_i(jF/2 + F - 1) \right]^{T}$ with $j=0,1,\dots,\Theta-1$. The energy value of the $i$-th frequency channel in the Gammatonegram at the $j$-th time instant is:
\begin{equation}
\mathcal{X}_{i,j} = \sqrt{ \frac{1}{F} \sum_{k = 0}^{F-1} \left[ \tilde{x}_i \left( j \frac{F}{2} + k \right) \right]^2 }.
\end{equation}
In Figure~\ref{fig:conf-prototype}, we show the Gammatonegram representation of a sample scream sound. It is similar to the spectrogram, with the substantial difference that the frequency axis has a logarithmic scale and the bandwidth of the band-pass filters increases linearly with the value of the central frequency.

\subsection{COPE features}
The configuration and application of COPE feature extractors involve a number of steps that we explain in the following of this section.
In the application phase, given the Gammatonegram representation of a sound, a COPE feature extractor responds strongly to patterns similar to the one used in the configuration step. 
It also accounts for some tolerance in the detection of the pattern of interest, so being robust to distortions due to noise or to varying SNR.

\subsubsection{Local energy peaks}
The energy peaks (local maxima) in a Gammatonegram $\mathcal{X}$ are highly robust to additive noise~\cite{wang03}. This property provides underlying robustness of the designed COPE features to variation of the SNR of the sounds of interest.
We consider that a point is a peak if it has higher energy than the points in its neighborhood. We suppress non-maxima points in the Gammatonegram and obtain an energy peak response map, as follows:

\begin{equation}
\mathcal{P_X}(t,f) = \max \limits_{\substack{t-\Delta t \leq t' \leq t + \Delta t \\ f-\Delta f \leq f' \leq f + \Delta f}} \mathcal{X}(t', f')
\end{equation}
where $t = 0, \dots, \Theta -1$ and $f = 0, \dots, \Gamma -1$.
The values $\Delta t$ and $\Delta f$ determine the size, in terms of time and frequency, of the neighborhood around a time-frequency point in which the local energy is evaluated (in this work we consider $8$-connected pixels).
We consider the arrangement (hereinafter constellation) of a set of such time-frequency points as a description of  the distribution of the energy of a particular sound.

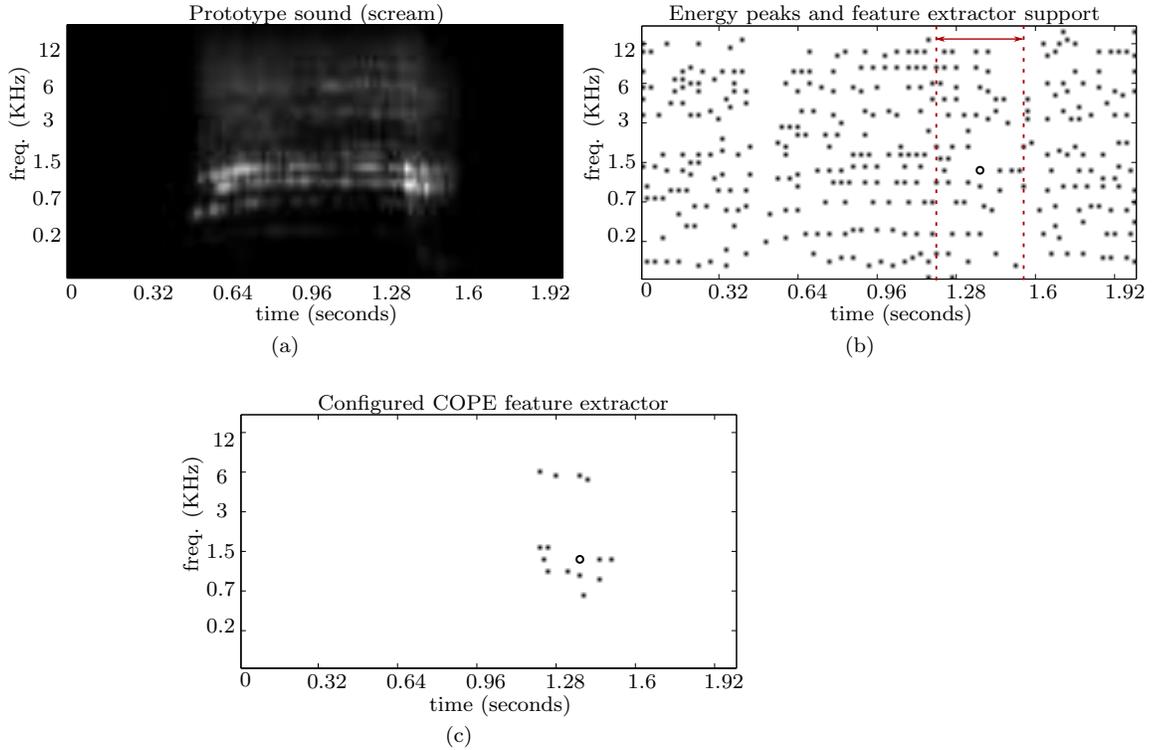
\begin{figure}[!t]
	\centering
	\footnotesize
   \subfloat[]{\label{fig:conf-prototype}
	\input{figures/method/prototype-paper.eps_tex}}
	~
	\subfloat[]{\label{fig:conf-peaks}
   \input{figures/method/allpeaks-paper.eps_tex}}
   	
   \subfloat[]{\label{fig:conf-selected}
   \input{figures/method/configuredpeaks-paper.eps_tex}}
   \caption{Example of configuration of a COPE feature extractor performed on the (a) Gammatonegram representation of a scream. The (b) energy peaks are extracted and a support (dashed lines) is chosen around a reference point (small circle). The (c) configured feature extractor is composed of only those points whose energy is higher than a fraction $t_1$ of the energy of the reference point.}
   \label{fig:examples}
\end{figure}

\subsubsection{Configuration of a COPE feature extractor}
\label{sec:method-conf}
Given the constellation of energy peaks of a sound and a reference point (in our case the point that correspond to the highest peak of energy), we determine the structure of a COPE feature extractor in an automatic configuration process. For the configuration one has to set the support size of the COPE feature extractor, i.e. the size of the time interval around the reference point in which to consider energy peaks. 

In Figure~\ref{fig:examples}, we show an example of the configuration process on the scream sound in Figure~\ref{fig:conf-prototype}. First, we find the position of the local energy peaks and select a reference point (small circle in Figure~\ref{fig:conf-peaks}) around which we define the support size of the feature extractor. The support is contained between the two dashed (red) lines in Figure~\ref{fig:conf-peaks}. 
We consider the positions of only those peaks that fall within the support of the feature extractor and whose energy is higher than a fraction $t_1$ of the highest peak of energy (Figure~\ref{fig:conf-selected}).
Every peak point $p_i$ is represented by a $3$-tuple $(\Delta t_i,f_i,e_i)$: $\Delta t_i$ is the temporal offset with respect to the reference point, $f_i$ is its corresponding frequency channel in the Gammatone filterbank and $e_i$ is the value of the energy contained in it. 

The configuration process results in a set of tuples that describe the constellation of energy peaks in the Gammatonegram image of a sound.
We denote by $S= \{(\Delta t_i,f_i,e_i)\mid i=1,\dots,L\}$ the set of $3$-tuples, where $L$ is the number of considered peaks within the support of the filter.

\subsubsection{Feature computation}
Given a Gammatonegram, we compute the response of a COPE feature extractor as a combination of its weighted and shifted energy peaks. We define the weighting and shifting of the $i$-th energy peak as :
\begin{equation}
s_i(t) = \max\limits_{t',f'}{ \left\{ \psi(t -\Delta  t_i - t', f_i -\Delta f_i -f') G_{\sigma'}(t', f' ) \right\} }
\label{eq:blurring}
\end{equation}
where $-3 \sigma' \leq t', f' \leq 3 \sigma'$.

The function $\psi(t,f)$ can be seen as a response map of the similarity between the detected energy peak in the input Gammatonegram and the corresponding one in the model. In this work we consider $\psi(t,f) = \mathcal{P_X}(t,f)$, so as to account only for the position and energy content of the peak points in the constellation. 
We weigth the response $\psi(t,f)$ with a Gaussian weighting function $G_{\sigma'}(\cdot,\cdot)$ that allows for some tolerance in the expected position of the peak points. This choice is supported by evidence in the auditory system that vibrations of the cochlea membrane due to a sound wave of a certain frequency excite neurons specifically tuned for that frequency and also neighbor neurons~\cite{PALMER19861}. 
The size of the tolerance region is determined by the standard deviation $\sigma'$ of the function $G_{\sigma'}$, which is a parameter and we set as $\sigma' = \sigma_0 / 2$ . 


The value of a COPE feature is computed with a sliding window that shifts on the Gammatonegram of the input sound. Formally, we define it as the geometric mean of the weighted and shifted energy peak responses in Eq.~\ref{eq:blurring}:
\begin{equation}
r(t) =  \left\lvert \left( \prod_{i=1}^{\lvert S \lvert} s_i(t)\right)^{1/\lvert S \lvert}\right\lvert_{t_2},
\end{equation}
where $t_2$ is a threshold value. Here, we set $t_2=0$, so to not suppress any response. 
The value of a COPE feature for a sound in an interval delimited by $[T_1,T_2]$ is given by max-pooling of the response $r(t)$ with $T_1 \leq t \leq T_2$:
\begin{equation}
\hat{r} = \max\limits_{t \in [T_1,T_2]} r_i(t)
\end{equation}

\subsection{COPE feature vector}
We configure a set of COPE feature extractors on $K$ training audio samples from different classes. For a given interval of sounds $[T_1,T_2]$, we then construct a feature vector  as follows:
\begin{equation}
\mathbf{v}_{[T_1, T_2]} = \left[ \hat{r}_1, \hat{r}_2, \hdots, \hat{r}_K \right].
\end{equation}

\subsection{Classifier}
We use the COPE feature vectors to train a classifier, which is able to assign the input sound to one of the $M$ classes of interest. The COPE feature vectors are not dependent on a specific classier and thus one can employ them together with any multi-class classifier.

In this work, we employ a multi-class SVM classifier, designed according to a \emph{one-vs-all} scheme, in which $M$ binary SVM classifiers (where $M$ corresponds to the number of classes) are trained to recognize samples from the classes of interest. We use linear SVMs with soft-margin as they provide already satisfactory results and are easy to train. We set the hyperparameter $c=1$ for the training of each SVM, which indicates the trade-off between training error and size of the classification margin while training the SVM classifier (see~\cite{CortesSVM1995} for reference). We train the $i$-th SVM ($i=0, \dots, M-1$) by using as positive examples those of the class $C_i$ and as negative samples those of all the remaining classes.
In this scheme, the training of each SVM classifier is an unbalanced problem, as the cardinality of the samples from the negative class $\lvert N \lvert$ outnumbers that of the samples from the positive class  $\lvert P \lvert$. We thus employ an implementation of the SVM algorithm that includes a cost-factor $J = \lvert N \lvert / \lvert P \lvert$ by which training errors on positive examples outweight errors on negative examples\footnote{Available in the SVMlight library - http://svmlight.joachims.org/}~\cite{Morik99}. In this way, the training errors for the positive and negative examples have the same influence in the overall optimization.

During the test phase, each SVM classifier assigns a score $m_i$ to the given sample under test (i.e. a COPE feature vector that represents the sound to classify). We analyze the SVM scores $m_i$ and assign to the test vector the class that corresponds to the SVM that gives the highest classification score. We assign the sample under test to the reject class $C_0$ (background sound) in case all the scores are negative. We formally define the classification rule as:

\begin{equation}
C = \begin{cases} C_0, &\quad if~m_i < 0 \qquad \forall i=0, \dots, M-1 \\
	\displaystyle\arg\max_{i} m_i,  &\quad else.
\end{cases}
\label{eq:combrule}
\end{equation}


\section{Data sets}
\label{sec:dataset}
We carried out experiments on four public data sets, namely the MIVIA audio events~\cite{BoawPRL2015}, MIVIA road events~\cite{ITS2015}, ESC-10~\cite{Piczak15} and TU-Dortmund~\cite{Plinge2014} data sets. 

\subsection{MIVIA audio events}
\label{sec:dataset-1}
Typical sounds of interest for intelligent surveillance applications are glass breakings, gun shots and screams. In the MIVIA audio events data set, such sounds are superimposed to various background sounds and have different SNRs ($\{5, 10, \dots,30\} dB$). 
This simulates the occurrence of sounds  in different environments and at various distances from the microphone.
We extended the  data set by including cases in which the energy of the sounds of interest is equal or lower than the one of the background sound, so having null or negative SNR. 
Thus, adopting the same procedure described in~\cite{BoawPRL2015}, we created two versions of the audio events at $0 dB$ and $-5 dB$ SNR.
The final data set\footnote{The data set is publicly available at the url \emph{http://www.gitlab.com/nicstrisc/COPE}} contains a total of $8000$ events for each class, divided into $5600$ events for training and $2400$ events for testing equally distributed over the considered values of SNR. The audio clips are PCM sampled at $32$KHz with a resolution of $16$ bits per sample. Hereinafter we refer at glass breaking with \emph{GB}, at gun shots with \emph{GS} and at screams with \emph{S}. We indicate the background sound with \emph{BN}. 
In Table~\ref{tab:event-desc}, we report the details of the composition of the extended data set.

\begin{table}[!t]%
	\footnotesize
	\renewcommand{\arraystretch}{0.9}
	\centering
		\caption{Details of the composition of the MIVIA audio events data set. The total duration of the sounds is expressed in seconds.}
    		\begin{tabular}{c|cc|cc}
			\multicolumn{5}{c}{\bfseries MIVIA audio events data set} \\ \hline \hline
			~ & \multicolumn{2}{c|}{\bfseries Training set} & \multicolumn{2}{c}{\bfseries Test set} \\ \hline
       		~ &	\textbf{\#Events} & \textbf{Duration (s)} & \textbf{\#Events} & \textbf{Duration (s)} \\ \hline
       		\textbf{BN} & - & 77828.8 & - & 33382.4 \\
			\textbf{GB} & 5600 & 8033.1 & 2400 & 3415.6 \\
			\textbf{GS} & 5600 & 2511.5 & 2400 & 991.3 \\
			\textbf{S}  & 5600 & 7318.4 & 2400 & 3260.5 \\
			\noalign{\hrule height 1.5pt}
		\end{tabular}
	\label{tab:event-desc}
\end{table}

\subsection{MIVIA road events}
\label{sec:dataset-2}
The MIVIA road events data set contains car crash and tire skidding events mixed with typical road background sounds such as traffic jam, passing vehicles, crowds, etc. 
A total of $400$ sound events ($200$ car crashes and $200$ tire skiddings) are superimposed to various road  sounds ranging from very quiet (e.g. in country roads) to highly noisy traffic jams (e.g. in the center of a big city) and highways.
The sounds of interest are distributed over $57$ audio clips of about one minute each, which are organized into four independent folds (in each fold $50$ events per class are present) for cross-validation experiments. The audio signals are sampled at $32$KHz with a resolution of $16$ bits per PCM sample. In the rest of the paper, we refer at car crash with \emph{CC} and at tire skidding with \emph{TS}. 

%

\subsection{ESC-10}
The ESC-10 data set is composed of $400$ sounds divided in ten classes (\emph{dog bark, rain, sea waves, baby cry, clock tick, sneeze, helicopter, chainsaw, rooster, fire crackling}), each of them containing $40$ samples. The sounds are sampled at $44.1$ KHz with a bit rate of $192$ kbit/s and their total duration is about $33$ minutes. The data set is organized in five independent folds. The average classification accuracy achieved by human listeners is $95.7\%$.

\subsection{TU Dortmund}
The TU Dortmund data set was recorded in a smart room with a microphone embedded on a table. The data set is composed of sounds from eleven classes (\emph{chair, cup, door, keyboard, laptop keys, paper sheets, pouring, rolling, silence, speech, steps}), divided in a training and a test sets. The sounds of interest are sampled at $48$ Khz and are mixed with the background sound of the smart room. A ground truth with the start and end points of the sounds is provided. 
We constructed a second observer ground truth, which contains a finer grain manual segmentation of the events. 


\section{Experiments}
\label{sec:experiments}

\subsection{Performance evaluation}
\label{sec:experiments-eval}
For the MIVIA audio events and the MIVIA road events data sets we adopted the experimental protocol defined in~\cite{BoawPRL2015}. The performance evaluation is based on the use of a time window of $T_w$ seconds that forward shifts on the audio signal by $\Delta T_w$ seconds.  An event is considered correctly detected if it is detected in at least one of the time windows that overlap with it.
Besides the recognition rate and confusion matrix, we consider two types of error that are important for performance evaluation: the detection of events of interest when only background sound is present (false positive) and the case when an event of interest occurs but it is not detected (missed detection). In case a false positive is detected in two consecutive time windows, only one error is counted.
We measured the performance of the proposed method by computing the recognition rate (RR), false positive rate (FPR), error rate (ER) and miss detection rate (MDR). 
Moreover, in addition to the receiver operating characteristic (ROC) curve and in order to assess the overall performance of the proposed method we compute the Detection Error Trade-off (DET) curve. It is a plot of the trade-off between the false positive rate and the miss detection rate and gives an insight of the performance of a classifier in terms of its errors. In contrast to the ROC curve, in the DET curve the axis are logarithmic in order to highlight differences between classifiers in the critical operating region. The closer the curve to the point $(0,0)$, the better the performance of the system. 

For the ESC-10 and TU Dortmund data sets, we evaluate the performance for the classification of isolated audio events. This type of evaluation is done according to the structure of these data sets and to make possible a comparison with the results achieved by other approaches. We compute the average recognition rate (RR) and the F-Measure $F=2 Re Pr/(Re+Pr)$, where $Pr=TP/(TP+FP)$ is the precision and $Re=TP/(TP+FN)$ is the recall. $TP$, $FP$ and $FN$ are the number of true positive, false positive and false negative classifications, respectively. 
In the case of the MIVIA road events and the ESC-10 data sets, we perform cross-validation experiments. 




\subsection{Results}
\label{sec:results}
In Table~\ref{tab:mat-full-cosfire}, we report the classification matrix that we obtained on the extended version of the MIVIA audio events data set. 
The average recognition rate for the three classes is $90.7\%$, while the miss detection rate and the error rate are $3.7\%$ and $5.6\%$, respectively.
We obtained an FPR equal to $7.1\%$, of which $1.25\%$ are glass breakings, $2.74\%$ are gun shots and $3.11$ are screams.

\begin{table}[!t]%
	\renewcommand{\arraystretch}{0.9}
	\centering
		\footnotesize
	\caption{Classification matrix obtained by the proposed method on the extended MIVIA audio events data set. GB, GS and S indicate the classes in the data set (see Section~\ref{sec:dataset-1}), while MDR is the miss detection rate.}
    		\begin{tabular}{C{0.9cm}|C{1cm}|ccc|c}
			\multicolumn{6}{c}{\bfseries Results - MIVIA audio events data set} \\
			\noalign{\hrule height 1.5pt}
			\multicolumn{2}{c}{~} & \multicolumn{3}{c|}{\textbf{Detected class}} & ~ \\ \cline{3-6}
    			\multicolumn{2}{c}{~} & \textbf{GB} & \textbf{GS} & \textbf{S} & \textbf{MDR}\\ \hline
			\multirow{3}{*}{\newlinecell{\textbf{True}\\ \textbf{class}}}  
				& \textbf{GB} & $95.33\%$ & $2.13\%$ & $1.25\%$ & $1.29\%$ \\
			~ & \textbf{GS} & $4.33\%$ & $89.25\%$ & $2.58\%$ & $3.83\%$ \\
			~ & \textbf{S}  & $1.5\%$ & $4.92\%$ & $87.79\%$ & $5.79\%$  \\
			
			\noalign{\hrule height 1.5pt}
		\end{tabular}
	\label{tab:mat-full-cosfire}
\end{table}

In Table~\ref{tab:mat-road-cosfire} we report the classification matrix achieved by the proposed approach on the MIVIA road events data set. The average RR is $94\%$ with a standard deviation of $4.32\%$, while the average FPR is $3.94\%$ with a standard deviation of $1.82\%$. The results are in line with the ones achieved on the MIVIA audio events data set. The low standard deviation of the recognition rate is indicative of good generalization capabilities.


The proposed method shows high performance on the ESC-10 and TU Dortmund data sets, which both contain a larger number of classes than in the MIVIA data sets, but with a lower number of samples per class. We achieved $RR=81.25\%~~(5.38\%),  Pr=0.8263~~(0.053), Re=0.8125~~(0.054), F=0.8048~~(0.059)$ on the ESC-10 data set (the standard deviation of each measure is in brackets). On the TU Dortmund data set we achieved $RR=94.27 \%, Pr= 0.9479, Re= 0.9519$ and $F=0.9469$. In the following, we compare the achieved results with the ones reported in other works.

\begin{table}[!t]%
	\renewcommand{\arraystretch}{0.9}
	\centering
		\footnotesize
	\caption{Average results obtained by the proposed method on the MIVIA road events data set. CC and TS are acronyms for the classes in the data set (see Section~\ref{sec:dataset-2}).}
    		\begin{tabular}{C{1.2cm}|C{1.2cm}|C{1.2cm}C{1.2cm}|C{1.2cm}}
			\multicolumn{5}{c}{\bfseries Results - MIVIA road events} \\
			\noalign{\hrule height 1.5pt}
			\multicolumn{2}{c}{~} & \multicolumn{2}{c|}{\textbf{Guessed class}} & ~ \\ \cline{3-5}
    			\multicolumn{2}{c}{~} & \textbf{CC} & \textbf{TS} & \textbf{MDR}\\ \hline
			\multirow{2}{*}{{\newlinecell{\textbf{True}\\ \textbf{class}}}}  
				& \textbf{CC} & $92\%$ & $2\%$ & $6\%$ \\
			~ & \textbf{TS} & $0.5\%$ & $96\%$ & $3.5\%$ \\
			\noalign{\hrule height 1.5pt}
		\end{tabular}
	\label{tab:mat-road-cosfire}
\end{table}


\subsection{Results comparison}

\begin{table*}[!t]
  \renewcommand{\arraystretch}{0.9}
  \centering
  	\footnotesize
  	\caption{Comparison of the results with the ones of existing approaches on the MIVIA audio events data set. RR, MDR, ER and FPR refer to the metrics described in Section~\ref{sec:experiments-eval}.}
    \begin{tabular}{c|C{10mm}C{10mm}C{10mm}C{10mm}}
    	\multicolumn{5}{c}{\bfseries Result comparison - MIVIA audio events data set} \\
    	\noalign{\hrule height 1.5pt}  
	\textbf{Method} & \textbf{RR} & \textbf{MDR}  & \textbf{ER} & \textbf{FPR} \\ \hline \hline
	\multicolumn{5}{c}{Test with $SNR > 0$} \\ \hline 
	COPE & $96\%$ & $\bm{3.1\%}$ & $\bm{0.9\%}$	& $4.3\%$ \\
	$bof_{h}$~\cite{BoawPRL2015} &$84.8\%$ & $12.5\%$ & $2.7\%$ & $2.1\%$ \\
	$bof_{s}$~\cite{BoawPRL2015} & $86.7\%$ & $10.7\%$ & $2.6\%$ & $3.1\%$  \\ 
	Gammatone~\cite{Saggese16} & $88.6\%$ & $9.65\%$ & $1.4\%$ & $\bm{1.4}\%$  \\
	UDWT~\cite{Saggese16} & $77.81\%$ & $10.65\%$ & $11.54\%$ & $6.6\%$  \\
	SoundNet~\cite{SoundNet} & $93.33\%$ & $0.67\%$ & $6\%$ & $22.34\%$  \\
	HRNN~\cite{Colangelo17} & $\bm{96.55}\%$ & $-$ & $-$ & $-$  \\\hline \hline
	 
	 \multicolumn{5}{c}{Test with $SNR > 0$ and $SNR \le 0$} \\ \hline 
	COPE & $\bm{91.7\%}$ & $\bm{2.61\%}$ & $\bm{5.68\%}$	& $9.2\%$ \\
	$bof_{h}$~\cite{BoawPRL2015} &  $56.07\%$ & $36.43\%$ & $7.5\%$ 	& $\bm{5.3\%}$\\
	$bof_{s}$~\cite{BoawPRL2015} & $59.11\%$ & $32.97\%$ & $7.92\%$ & $\bm{5.3}\%$ \\ 
	SoundNet~\cite{SoundNet} & $84.13\%$ & $4\%$ & $11.88\%$ & $25.9\%$  \\
    \noalign{\hrule height 1.5pt}
    \end{tabular}%
  \label{tab:MIVIA-comparison}
\end{table*}%

In Table~\ref{tab:MIVIA-comparison}, we report the results that we achieved on the MIVIA audio event data set, compared with the ones of existing methods. In the upper part of the table we compare the results achieved by considering the classification of sound events with positive SNR only. In the lower part of the Table, we report the results achieved by including also sound events with negative and null SNR in the evaluation.

It is important to clarify that the methods described in~\cite{BoawPRL2015, Saggese16} employ the same multi-class \emph{one-vs-all} linear SVM classifiers of this work. The results that we report using SoundNet features~\cite{SoundNet} were obtained by using the features computed at the last convolutional layer of the SoundNet network in combination with the same classifier of this work.

SoundNet features obtained comparable recognition rate to the one achieved by the proposed approach, but a considerably higher FPR.
The recognition rate achieved by the Hierarchical Recurrent Neural Network classifier (HRNN) proposed in~\cite{Colangelo17} is slightly higher than the ones we obtained, though the HRNN-based approach has more complex design and training procedure, and a different classifier than SVM. The values of MDR, ER and FPR are not reported in~\cite{Colangelo17}. 

The performance of the proposed method demonstrated high robustness of the COPE feature extractors w.r.t. variations of the SNR. Conversely, for the methods proposed in~\cite{BoawPRL2015}, the performance of the classification systems strongly depend on the SNR of the training sound events.
When sounds with only positive SNR are used for training, the recognition rate achieved by the proposed method is almost $10\%$ higher than the one obtained by the approaches proposed in~\cite{BoawPRL2015,Saggese16}. The performance results of the latter methods decrease strongly (recognition rate more than $30\%$ lower than the one of the proposed method) when sounds with negative SNR are included in the model. We provide an extensive analysis of robustness to variations of SNR in Section~\ref{sec:snr}.

\begin{table}[!t]%
	\renewcommand{\arraystretch}{0.9}
		\footnotesize
	\centering
	\caption{Comparison of the results achieved on the MIVIA roads events data set with respect to the methods proposed in~\cite{ITS2015,Carletti13}. RR, MDR, ER and FPR refer to the evaluation metrics described in Section~\ref{sec:experiments-eval}.}
    		\begin{tabular}{c|C{11mm}C{11mm}C{11mm}C{11mm}}
			\multicolumn{5}{c}{\bfseries Comparison of results on MIVIA road events data set} \\ \noalign{\hrule height 1.5pt} 
       		~ &	\bfseries RR & \bfseries MR & \bfseries ER & \bfseries FPR \\ \hline \hline
       		COPE & $94\%$ & $4.75\%$ & $1.25\%$ & $3.95\%$ \\
       		$\sigma$ & $4.32$ & $4.92$ & $1.26$ & $1.82$ \\ \hline
       		$bof_{BARK}$~\cite{ITS2015} & $80.25\%$ & $21.75\%$ & $3.25\%$ & $10.96\%$ \\
       		$\sigma$ & $7.75$ & $8.96$ & $2.5$ & $8.43$ \\ \hline
       		$bof_{MFCC}$~\cite{ITS2015} & $80.25\%$ & $19\%$ & $0.75\%$ & $7.69\%$ \\
       		$\sigma$ & $11.64$ & $11.63$ & $0.96$ & $5.92$ \\ \hline
       		$bof$ \cite{Carletti13, ITS2015} & $82\%$ & $17.75\%$ & $0.25\%$ & $2.85\%$ \\
       		$\sigma$ & $7.79$ & $8.06$ & $1$ & $2.52$ \\ \hline
			\noalign{\hrule height 1.5pt}
		\end{tabular}
	\label{tab:compare-roads}
\end{table}

In Table~\ref{tab:compare-roads}, we compare the results we obtained on the MIVIA road events data set with the ones reported in~\cite{ITS2015}, where different sets of audio features (BARK, MFCC and a combination of temporal and spectral features) have been employed as short-time descriptors of  sounds. 
We obtained an average recognition rate ($94\%$) that is  more than $10\%$ higher than the ones achieved by existing methods, with a lower standard deviation. 

\begin{figure}[!t]
	\centering
	\scriptsize
	\subfloat[]{\label{fig:det-events}
   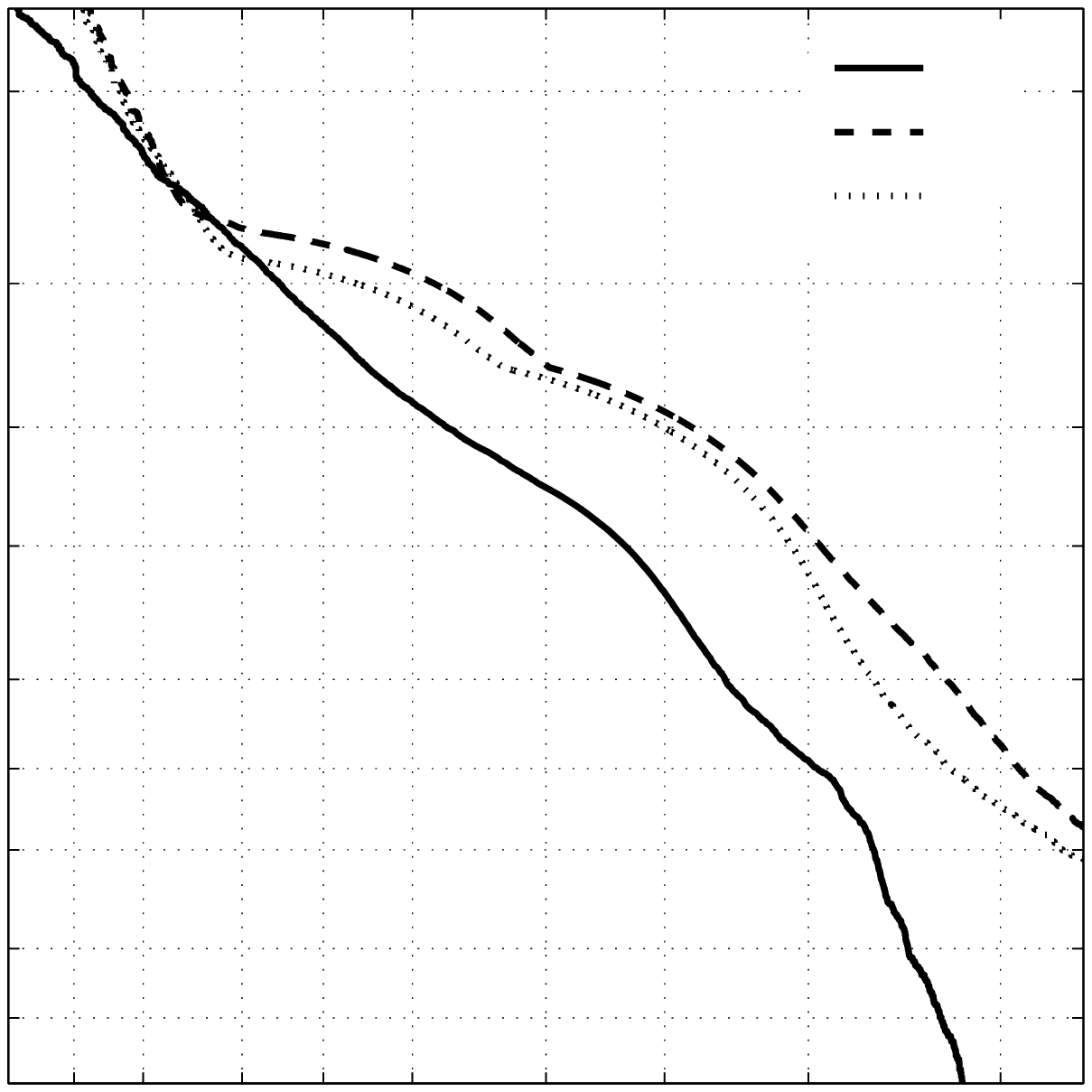}
~
   \subfloat[]{\label{fig:det-roads}
   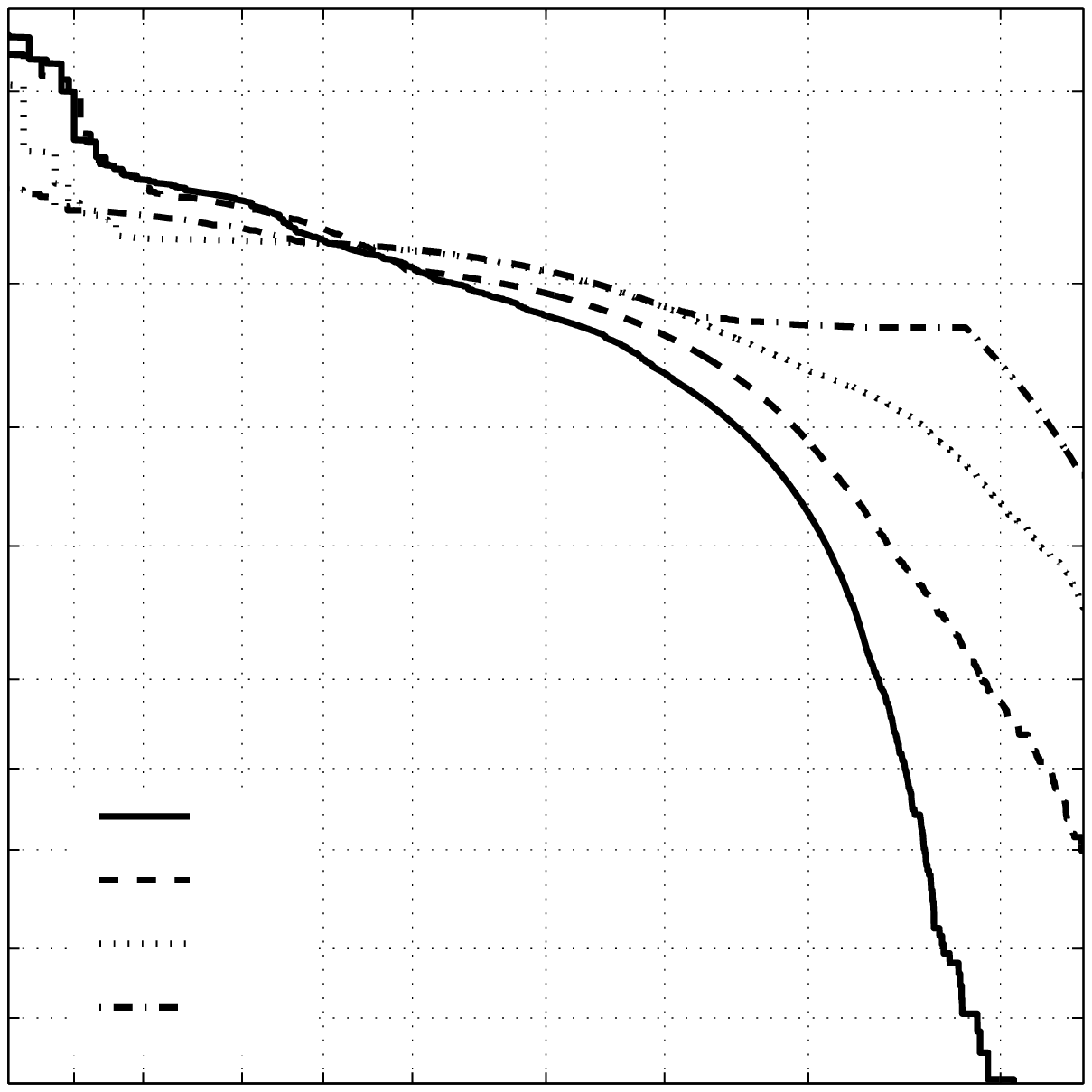}
   \caption{Detection Error Trade-off curves achieved by the proposed method (solid line) compared to the curves achieved by existing methods (dashed lines) on the (a) MIVIA audio events  and (b) MIVIA road events data sets. (Notice the logarithmic scales.)}
   \label{fig:detcurves}
\end{figure}

In Figure~\ref{fig:det-events} and~\ref{fig:det-roads}, we plot the DET curves obtained by our method (solid lines) on the MIVIA audio events and MIVIA road events data sets, respectively, and those of the methods proposed in~\cite{BoawPRL2015} and~\cite{ITS2015} (dashed lines). 
The curve of our method is closer to the point $(0,0)$ than the ones of other approaches, so confirming higher performance with respect to existing methods on the concerned data sets.

We compare the results that we achieved on the \mbox{ESC-10} data set with the ones reported by existing approaches in Table~\ref{tab:ESC10comparison}. The sign `$-$' indicates that the concerned value is not reported in published papers. The highest recognition rate is achieved by SoundNet~\cite{SoundNet}, which is a deep neural network trained on a very large data set of audio-visual correspondences. Approaches based on CNNs (\cite{Piczak15a,SoundNet,Hertel16,Medhat2017}) are trained with data augmentation techniques and generally perform better than the proposed approach on the ESC-10 data set, which is instead trained only on the original sounds in the ESC-10 data set.


\begin{table}[!t]%
	\renewcommand{\arraystretch}{0.9}
	\centering
		\footnotesize
	\caption{Comparison of the results on the ESC-10 data set. In brackets we report the standard deviation of the average performance metrics.  RR, MDR, ER and FPR refer to the evaluation metrics described in Section~\ref{sec:experiments-eval}.}
    		\begin{tabular}{c|C{18mm}C{18mm}}
			\multicolumn{3}{c}{\bfseries Result comparison on ESC-10 data set } \\
			\noalign{\hrule height 1.5pt}
    			\bfseries Method & \textbf{RR} &  \textbf{F}\\ \hline \hline
			 COPE  & $81.25\%~(5.38)$ & $ 0.81~ (0.06)$\\ 
			 Baseline~\cite{Piczak15}  & $66.74\%~(6.11)$ & $-$ \\ 
		     Random Forest~\cite{Piczak15} & $72.75\%~(8.68)$ & $0.72 ~(0.09)$ \\ 
			 Piczak CNN~\cite{Piczak15a} & $80.25\% ~(5.48)$  & $0.80 ~(0.06)$ \\ 
			 Conv. Autoenc.~\cite{SoundNet} & $74.3\%$ &  $-$ \\ 
			 Hertel CNN~\cite{Hertel16} & $89.9\%$ & $-$ \\ 
			 SoundNet~\cite{SoundNet} & $92.2\%$ & $-$\\ 
			 MCLNN.~\cite{Medhat2017} & $85.5\%$ & $-$ \\ 
			\noalign{\hrule height 1.5pt}
		\end{tabular}
	\label{tab:ESC10comparison}
\end{table}

In Table~\ref{tab:compTUdortmund}, we  report the results that we achieved on the TU Dortmund data set together with those reported in~\cite{Plinge2014}, where a classifier based on bag of features was proposed.
Besides the traditional bag of features (BoF) scheme, the authors proposed a pyramidal approach (P-BoF) and the use of super-frames (BoSF) for embedding temporal information about the sequence of features. The results in Table~\ref{tab:compTUdortmund} are computed according to the ground truth that we constructed based on a fine segmentation of sounds of interest and that we made publicly available.
It is worth noting that the performance results of our method refer to the classification of sound events. For the methods proposed in~\cite{Plinge2014} the evaluation is performed by considering the classification of sound frames.

\begin{table}[!t]%
	\renewcommand{\arraystretch}{0.9}
	\centering
		\footnotesize
	\caption{Comparison of the results on the TU Dortmund data set. The results were computed with respect to the second observer ground truth that we constructed. RR, MDR, ER and FPR refer to the evaluation metrics described in Section~\ref{sec:experiments-eval}.}
    		\begin{tabular}{c|C{12mm}C{12mm}C{12mm}C{12mm}}
			\multicolumn{5}{c}{\bfseries Result comparison on TU Dortmund data set } \\
			\noalign{\hrule height 1.5pt}
    			~ & \textbf{RR} &  \textbf{Pr} &  \textbf{Re} &  \textbf{F}\\ \hline \hline
			\bfseries COPE & $\bm{94.27\%}$ & $\bm{94.79 \%}$ & $\bm{95.19 \%}$ & $\bm{94.69\%}$\\ 
			\bfseries BoF~\cite{Plinge2014}  & $90.05\%$ & $ 92.39 \%$ & $88.82 \%$ & $90.57\%$ \\ 
		    \bfseries P-BoF~\cite{Plinge2014} & $89.94\%$ & $ 92.24 \%$ & $88.67 \%$ & $90.42\%$ \\ 
			\bfseries BoSF~\cite{Plinge2014} & $90.31\%$ & $ 92.73 \%$ & $88.13 \%$ & $90\%$ \\ 
			\noalign{\hrule height 1.5pt}
		\end{tabular}
	\label{tab:compTUdortmund}
\end{table}

\subsection{Robustness to background noise and SNR variations}
\label{sec:snr}

\begin{table*}[!b]
  \renewcommand{\arraystretch}{0.9}
  \centering
  	\footnotesize
  	\caption{Analysis and comparison of stability of results w.r.t. varying value of SNR of the events of interest. Details on the training schemes T1 and T2 are provided in Section~\ref{sec:snr}. RR, MDR, ER and FPR refer to the evaluation metrics described in Section~\ref{sec:experiments-eval}.}
    \begin{tabular}{C{8mm}|C{12mm}|C{10mm}C{10mm}C{10mm}C{10mm}|C{10mm}C{10mm}C{10mm}C{10mm}}
    	\multicolumn{10}{c}{\bfseries Comparison of results on the MIVIA audio events data set} \\ \noalign{\hrule height 1.5pt} 
    \multicolumn{2}{c|}{~} & \multicolumn{4}{c|}{\textbf{Training T1}} & \multicolumn{4}{c}{\textbf{Training T2}}\\ \hline \hline
    
   \bfseries Test & \textbf{Method} & \textbf{RR} & \textbf{MDR}  & \textbf{ER} & \textbf{FPR} & \textbf{RR} & \textbf{MDR}  & \textbf{ER} & \textbf{FPR}\\  \hline
   \multirow{2}{*}{\rotatebox{90}{\hspace{-7mm}\bfseries all SNR}} & COPE  & $\bm{91.7\%}$ & $\bm{2.61\%}$ & $\bm{5.68\%}$	& $9.2\%$ & $\bm{90.7\%}$ & $\bm{3.7\%}$ & $\bm{5.6\%}$ & $7.2\%$ \\
   ~&  $bof_{h}$~\cite{BoawPRL2015}  & $76.4\%$ & $11.64\%$ & $11.96\%$ & $\bm{5.9\%}$ & $56.07\%$ & $36.43\%$ & $7.5\%$ 	& $\bm{5.3\%}$ \\
   ~&  $bof_{s}$~\cite{BoawPRL2015} & $77.81\%$ & $10.65\%$ & $11.54\%$ & $6.6\%$ & $59.11\%$ & $32.97\%$ & $7.92\%$ & $\bm{5.3}\%$ \\ \hline 
   \multirow{2}{*}{\rotatebox{90}{\hspace{-7mm}\bfseries SNR$>$0}} & COPE & $96\%$ & $3.1\%$ & $0.9\%$ & $4.3\%$ & $95.2\%$ & $4\%$ & $0.8\%$ & $2.2\%$ \\ 
    ~& $bof_{h}$~\cite{BoawPRL2015} & $84.8\%$ & $12.5\%$ & $2.7\%$ & $2.1\%$ & $64.63\%$ & $31\%$ 	& $4.4\%$ & $4.2\%$ \\ 
    ~& $bof_{s}$~\cite{BoawPRL2015} & $86.7\%$ & $10.7\%$ & $2.6\%$ & $3.1\%$ & $68.74\%$ & $26.4\%$ & $4.9\%$ & $4.5\%$ \\ 
    \noalign{\hrule height 1.5pt}
    \end{tabular}%
  \label{tab:events-comparison}
\end{table*}%

We carried out a detailed analysis of the performance of the COPE feature extractors on sounds  with different levels of SNR. 
We trained the proposed classifier following two different schemes. For the first training scheme (that we refer at as T1) we included in the training process only sound events with positive SNR. For the second training scheme (that we refer at as T2) we trained the classifier with all the sounds in the data set, including those with null and negative SNR. In Table~\ref{tab:events-comparison}, we report the results that we achieved for the classification of sounds in the MIVIA audio event data set by training the system according to T1 and T2. We tested both trained models on the whole test set of the MIVIA audio event data set (including negative and null SNRs). The proposed method showed stronger robustness to changing SNR w.r.t. previously published approaches in~\cite{BoawPRL2015}, especially when samples with null and negative SNR are not included in the training process. This demonstrate high generalization capabilities of the proposed COPE feature extractors to sound events corrupted by high-energy noise.

\begin{figure}[!h]
	\centering
	\small
   \input{figures/results/rocversions-paper.eps_tex}
   \vspace{0mm}
   \caption{ROC curves obtained by the proposed method on the MIVIA audio events data set at different SNR values ($\{ -5, 0, 5, \dots , 30\}dB$). The arrow indicates increasing values of SNR.}
   \label{fig:rocsnr}
\end{figure}
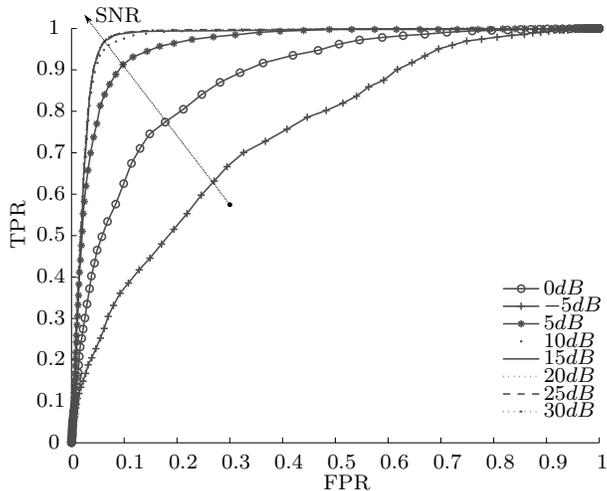

In Figure~\ref{fig:rocsnr}, we plot the ROC curves relative to the performance achieved at the different levels of SNR of the sounds of interest contained in the MIVIA audio event data set. We observed substantial stability of performance when the sounds of interest have positive (also very low) SNR. The  high robustness of the COPE feature extractors with respect to variations of the SNR is attributable to the use of the local energy peaks extracted from the Gammatonegram, which are robust to additive noise. 
The slightly lower results at negative SNR are mainly due to the changes of the expected energy peak locations caused by  high energy of the background sounds. In such cases, most of the wrong classifications are due to errors rather than to  miss detection of sounds of interest. 


\subsection{Sensitivity analysis}
We analyzed the sensitivity of the COPE feature extractors with respect to the parameter $\sigma_0$ which regulates the degree of tolerance to changes of the sounds of interest due to background noise or distortion.
We used a version of the MIVIA audio events data set specifically built for cross-validation experiments. The data set was released in~\cite{BoawPRL2015}, employing the same procedure used for the MIVIA audio events data set, ensuring statistical independence and high variability among folds. The sound events were divided into $k = 5$ independent folds, each of them containing $200$ events of interest per class (times $8$ versions of the SNR, as in the original MIVIA events data set). In our analysis, we estimated the variance of the generalization error using the Nadeau-Bengio variance estimator~\cite{Nadeau2003}, which takes into account the variability of the training and test sets used in cross-validation experiments.

For the configuration of a COPE feature extractor, the user has to choose the size of its support, i.e. the length of the time interval around the reference point in which  energy peaks are considered for the configuration (see Section~\ref{sec:method-conf}). 
We experimentally observed that different sizes of the support, namely $s_t = \{200, 300, 400\}$ ms, do not significantly influence the performance of the proposed system on the MIVIA data sets. 
We report results achieved with a support  of $200$ ms, which involves a limited number of energy peaks in the configuration of the feature extractors. 
One could however choose $s_t = 400$ ms, achieving similar performance to the case in which $s_t = 200$ ms. The drawback is the need of computing and combining the responses of a higher number of energy peaks, which increase the processing time of each feature extractor.

In Table~\ref{tab:sensitivity}, we report the generalization error (ER) and the false positive rate (FPR) as the parameter $\sigma_0$ varies. 
The performance of the proposed system is slightly sensitive to varying values of the parameter $\sigma_0$, mostly when they are kept very low. 
For higher values ($\sigma_0 = 3, 4, 5, 6$), the performance shows more stability. Higher tolerance for the detection of the energy peak positions determines stronger robustness to background noise. 
It is worth pointing out that too large values of tolerance might cause a loss in the selectivity and descriptive power of the COPE feature extractors and consequently a decrease of the classification performance.

\begin{table}[!t]%
	\renewcommand{\arraystretch}{0.9}
			\footnotesize
	\centering
	\caption{Sensitivity of the COPE feature extractors  to various values the parameter $\sigma_0$. Higher the value of $\sigma_0$, larger the tolerance of the feature extractor to variations of the pattern of interest. For the generalization error (ER) of cross-validation experiments, we report the value of the Nadeau-Bengio estimator of variance that takes into account the variability in the training and test sets~\cite{Nadeau2003}}.
    		\begin{tabular}{c|C{10mm}C{10mm}C{10mm}C{10mm}|C{10mm}C{10mm}C{10mm}C{10mm}}
			\multicolumn{9}{c}{\bfseries Sensitivity of COPE feature extractors } \\ \hline \hline
       		\multicolumn{1}{c|}{~} & \multicolumn{4}{c|}{\bfseries MIVIA audio events} & \multicolumn{4}{c}{\bfseries MIVIA road events}\\ \hline
			$\bm{\sigma_0}$ & $\bm{ER}$ & $\bm{\hat{\sigma}_{ER}}$ & $\bm{FPR}$ & $\bm{\sigma_{FPR}}$  & $\bm{ER}$ & $\bm{\hat}{\sigma_{ER}}$ & $\bm{FPR}$ & $\bm{\sigma_{FPR}}$ \\ \hline

		$\bm{1}$ & $35.98\%$ & $8.3$ & $19.21\%$ & $7.29$ & $28.5\%$ & $6.47$ & $17.09\%$ & $7.66$ \\
		$\bm{2}$ & $26.6\%$ & $3.92$ & $18.75\%$ & $3.16$ & $19\%$ & $3.87$ & $21.46\%$ & $11.14$\\
		$\bm{3}$ & $13.84\%$ & $1.41$ & $13.97\%$ & $2.85$ & $4.75\%$ & $4.41$ & $18.78\%$ & $12.5$\\
		$\bm{4}$ & $13.83\%$ & $2.1$ & $11.53\%$ & $2.4$ & $4.75\%$ & $3.08$ & $7.44\%$ & $4.39$\\
		$\bm{5}$ & $14.70\%$ & $2.22$ & $9.91\%$ & $2.2$ & $6\%$ & $3.05$ & $3.94\%$ & $1.82$\\
		$\bm{6}$ & $14.37\%$ & $3.04$ & $10.23\%$ & $2.1$ & $6.25\%$ & $3.39$ & $3.94\%$ & $1.13$\\ 
			\noalign{\hrule height 1.5pt}
		\end{tabular}
	\label{tab:sensitivity}
\end{table}

\section{Discussion}
\label{sec:discussion}
The high recognition capabilities of the proposed method are attributable to the trainable character and the versatility of the COPE feature extractors. 
The concept of trainable filters has been previously introduced for visual pattern recognition. COSFIRE filters were proposed for contour detection~\cite{AzzopardiPetkov2012}, keypoint and object detection~\cite{COSFIRE}, retinal vessel segmentation~\cite{AzzopardiStrisciuglio,StrisciuglioVIP15}, curvilinear structure delineation~\cite{StrisciuglioIWOBI17,StrisciuglioCAIP2017,StrisciuglioECCV18}, and action recognition~\cite{StrisciuglioAction2018}. 
In this work, we extended the concept of trainable feature extractors to sound recognition. 
It is noteworthy that the proposed COPE feature extractors do not relate with template matching techniques, which are sensitive to variations with respect to the reference pattern. 
The tolerance introduced in the application phase allows also for the detection of modified versions of the prototype pattern, mainly due to noise or distortion. 

An important advantage of using COPE feature extractors is the possibility of avoiding the process of feature engineering, which is a time-consuming task and requires substantial domain knowledge.
In traditional sound recognition approaches, hand-crafted features  (e.g. MFCC, spectral and temporal features, Wavelets and so on) are usually chosen and combined together to form a feature vector that describes particular characteristics of the audio signals. 
On the contrary, the automatic configuration of COPE feature extractors consists in learning data representations directly from the sounds of interest. 
Manual engineering of features is indeed not required.

Representation learning is typical of recent machine learning methods based on deep and convolutional neural networks, which require large amount of training data. When large data sets are not available, new synthetic data is generated by transformations of the original training data. To this concern, the COPE algorithm differs from deep and convolutional neural networks approaches, as it requires only one prototype pattern to configure a new feature. Moreover, the tolerance introduced in the application phase guarantees, to a certain extent, good generalization properties.
Because of their flexibility, COPE feature extractors can be thus employed in various sound processing applications such as music analysis~\cite{Newton11, Neocleous15} or audio fingerprinting~\cite{Cano05}, among others.

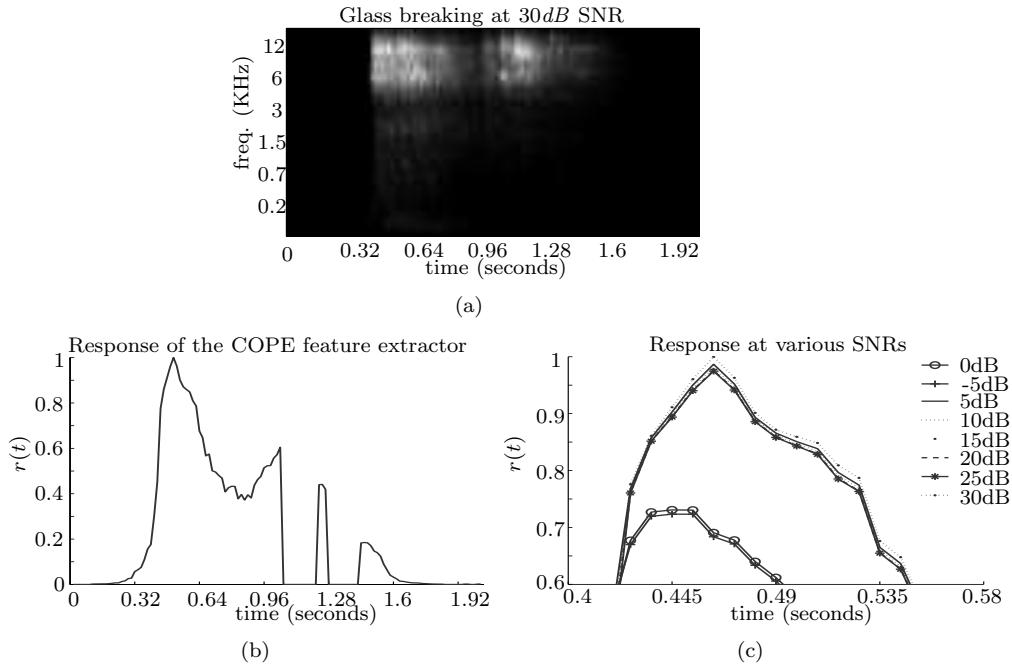
\begin{figure}[!t]
	\centering
	\footnotesize
	\subfloat[]{\label{fig:sound}
   \input{figures/discussion/sound-paper.eps_tex}}

   \subfloat[]{\label{fig:resp}
   \input{figures/discussion/resp-paper.eps_tex}}
~
   \subfloat[]{\label{fig:respdetail}
	\input{figures/discussion/respdetail-paper.eps_tex}}
   \caption{The (a) Gammatonegram of a prototype glass breaking sound used for the configuration of a COPE feature extractor. The (b) response $r(t)$ of the feature extractor computed on the sound in (a). 
The (c) time-zoomed (between $0.4$s and $0.58$s) response $r(t)$ at different SNRs ($\{-5,0,\dots,30\}$\textit{dB}). The response is stable for positive values of SNR and decreases for null or negative SNR values.  }
   \label{fig:response}
\end{figure}

The COPE feature extractors are robust to variations of the background noise and of the SNR of the sounds of interest. In Figure~\ref{fig:sound} we show the Gammatonegram of a glass breaking sound with SNR equal to $30$\textit{dB}. As an example, we configure a COPE feature extractor on this sound and compute its response $r(t)$ on the sound of Figure~\ref{fig:sound}, which we show in Figure~\ref{fig:resp}. One can observe that the response is maximum in the same point used as reference point in the configuration phase. The response is null when at least one of the expected energy peaks is not present. 
In Figure~\ref{fig:respdetail}, we show a time-zoomed detail of the response of the feature extractor computed on the same glass breaking event at different values of SNR (from $-5$\textit{dB} to $30$\textit{dB} in steps of $5$\textit{dB}). The response keeps stable for positive, also very low values of SNR and it slightly decreases for null or negative SNR values.
As demonstrated by the results that we reported in Section~\ref{sec:snr}, the stability of the response of COPE feature extractors and the high performance on sounds with different values of SNR.
The decrease of performance at null and negative SNR is due to the effect of  background sounds with  energy higher than that of the sounds of interest. It determines strong changes of the position of the energy peaks with respect to those determined in the configuration. To this concern, the effect of other functions $ \psi(f,t)$  in eq.~\ref{eq:blurring} to evaluate the energy peak similarity can be explored.

\section{Conclusions and outlook}
\label{sec:conclusion}
We proposed a novel method for feature extraction in audio signals based on trainable feature extractors, which we called COPE (that stands for Combination of Peaks of energy). We employed the COPE feature extractors in the task of environmental sound event detection and classification, and tested their robustness to variations of the SNR of the sounds of interest. The results that we achieved on four public data sets (recognition rate equal to $91.71 \%$ on the MIVIA audio events, $94 \%$ on the MIVIA road events, $81.25 \%$ on the ESC-10 and $94.27 \%$ on the TU Dortmund data sets) are higher than many existing approaches and demonstrate the effectiveness of the proposed method. 

The design of COPE feature extractors was based on neuro-physiological evidence of the mechanism that translates  sound pressure waves into neural stimuli from the cochlea membrane through the Inner Hair Cells (IHC) in the  auditory system of mammals. The proposed method can be extended by also including in its processing the implementation of a neuron response inhibition mechanism that prevents the short-time firing of those IHCs that have recently fired~\cite{Lopez-Poveda2006}. In this view, the computation of the energy peak map would need to account for the  energy distribution of the sounds of interest in each frequency band at a larger time scale, instead of performing a local analysis only. The extension of the COPE feature extractor with such inhibition phenomenon can further improve the robustness of the proposed method to changes of background noise and SNR, as only significant energy peaks are to be processed. 

Although the current implementation of the COPE feature extractor is rather efficient ($0.965$ seconds to compute a COPE feature vector of 200 elements for a signal of $3$ seconds, on a 2GHz dual core CPU), their computation can be further speeded-up. Parallelization approaches can be explored, which compute the value of COPE features or the local energy peak responses in separate threads. 
The construction of the COPE feature vector can also be optimized by including in the classification system only those filters that are relevant for the application at hand. A feature selection scheme based on the relevance of the feature values described can be employed~\cite{Strisciuglio2016}. 
The optimization of the number of configured feature extractors and the implementation of parallelization strategies can jointly contribute to the implementation of a real-time system for intelligent audio surveillance on edge embedded systems.


\appendix
\section{Biological motivation}
\label{sec:motivation}
The sound pressure waves that hit our ears are directed to the \emph{cochlea} membrane in the inner auditory system.
Different parts of the cochlea membrane vibrate according to the energy of the frequency components of the sound pressure waves~\cite{patterson1986auditory}. A bank of Gammatone filters was proposed as a model of the cochlea membrane, whose response over time forms a spectrogram-like image called Gammatonegram~\cite{Patterson1992}. 
The membrane vibrations stimulate firing of \emph{inner hair cells (IHC)}, which are neurons that lay behind the cochlea. The firing activity of IHCs stimulates various fibers of the auditory nerve over time. We consider the pattern of the IHC firing activity as a descriptor of the input sound.

Given a prototype sound, a COPE feature extractor models the pattern of points that describe the IHC firing activity. We consider the points of highest local energy in the Gammatonegram as the locations at which the IHCs fire, and the constellation that they form is a robust representation of the pattern of interest. Hence, a COPE feature extractor is  configured by modeling the constellation of the peak points of the Gammatonegram of a prototype sound. 



\section*{References}
\small
\bibliography{citations}

\end{document}

%% file: figures/method/schema-paper.eps_tex
\begingroup%
  \makeatletter%
  \providecommand\color[2][]{%
    \errmessage{(Inkscape) Color is used for the text in Inkscape, but the package 'color.sty' is not loaded}%
    \renewcommand\color[2][]{}%
  }%
  \providecommand\transparent[1]{%
    \errmessage{(Inkscape) Transparency is used (non-zero) for the text in Inkscape, but the package 'transparent.sty' is not loaded}%
    \renewcommand\transparent[1]{}%
  }%
  \providecommand\rotatebox[2]{#2}%
  \ifx\svgwidth\undefined%
    \setlength{\unitlength}{345bp}%
    \ifx\svgscale\undefined%
      \relax%
    \else%
      \setlength{\unitlength}{\unitlength * \real{\svgscale}}%
    \fi%
  \else%
    \setlength{\unitlength}{\svgwidth}%
  \fi%
  \global\let\svgwidth\undefined%
  \global\let\svgscale\undefined%
  \makeatother%
  \begin{picture}(1,0.40)%
    \put(0,0.025){\includegraphics[width=\unitlength]{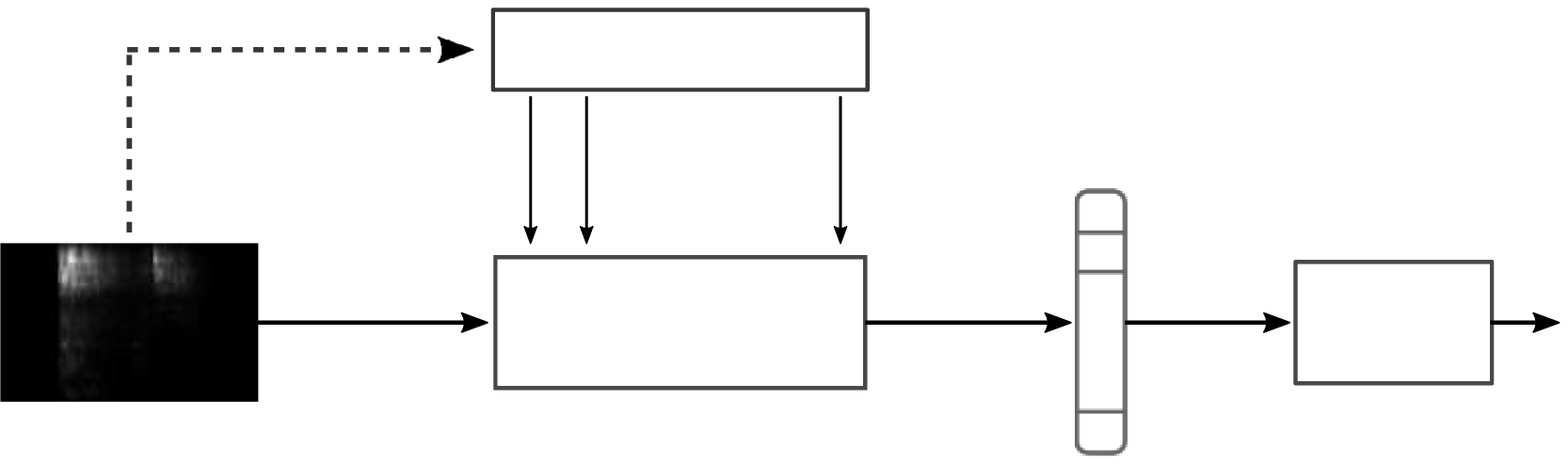}}%
    \put(0.355,0.285){\color[rgb]{0,0,0}\makebox(0,0)[lb]{\smash{COPE features}}}%
    \put(0.358,0.125){\color[rgb]{0,0,0}\makebox(0,0)[lb]{\smash{COPE feature}}}%
    \put(0.331,0.10217741){\color[rgb]{0,0,0}\makebox(0,0)[lb]{\smash{vector computation}}}%
    \put(0.838,0.125){\color[rgb]{0,0,0}\makebox(0,0)[lb]{\smash{Multi-class}}}%
    \put(0.867,0.10217741){\color[rgb]{0,0,0}\makebox(0,0)[lb]{\smash{SVM}}}%
    \put(0.98395225,0.13577507){\color[rgb]{0,0,0}\makebox(0,0)[lb]{\smash{C}}}%
    \put(0.45,0.2){\color[rgb]{0,0,0}\makebox(0,0)[lb]{\smash{...}}}%
    \put(0.7,0.125){\color[rgb]{0,0,0}\rotatebox{-90}{\makebox(0,0)[lb]{\smash{...}}}}%
    \put(0.018,0.3){\color[rgb]{0,0,0}\makebox(0,0)[lb]{\smash{configuration process}}}%
    \put(0.06762392,0.003){\color[rgb]{0,0,0}\makebox(0,0)[lb]{\smash{(a)}}}%
    \put(0.42,0.35){\color[rgb]{0,0,0}\makebox(0,0)[lb]{\smash{(b)}}}%
    \put(0.42,0.003){\color[rgb]{0,0,0}\makebox(0,0)[lb]{\smash{(c)}}}%
    \put(0.69,0.003){\color[rgb]{0,0,0}\makebox(0,0)[lb]{\smash{(d)}}}%
    \put(0.885,0.003){\color[rgb]{0,0,0}\makebox(0,0)[lb]{\smash{(e)}}}%
    \put(0.728,0.19){\color[rgb]{0,0,0}\makebox(0,0)[lb]{\smash{$\hat{r}_1$}}}%
    \put(0.728,0.16){\color[rgb]{0,0,0}\makebox(0,0)[lb]{\smash{$\hat{r}_2$}}}%
    \put(0.728,0.04){\color[rgb]{0,0,0}\makebox(0,0)[lb]{\smash{$\hat{r}_L$}}}%
  \end{picture}%
\endgroup%

%% file: figures/method/gammatone/impulse-low-paper.eps_tex
\begingroup%
  \makeatletter%
  \providecommand\color[2][]{%
    \errmessage{(Inkscape) Color is used for the text in Inkscape, but the package 'color.sty' is not loaded}%
    \renewcommand\color[2][]{}%
  }%
  \providecommand\transparent[1]{%
    \errmessage{(Inkscape) Transparency is used (non-zero) for the text in Inkscape, but the package 'transparent.sty' is not loaded}%
    \renewcommand\transparent[1]{}%
  }%
  \providecommand\rotatebox[2]{#2}%
  \ifx\svgwidth\undefined%
    \setlength{\unitlength}{180bp}%
    \ifx\svgscale\undefined%
      \relax%
    \else%
      \setlength{\unitlength}{\unitlength * \real{\svgscale}}%
    \fi%
  \else%
    \setlength{\unitlength}{\svgwidth}%
  \fi%
  \global\let\svgwidth\undefined%
  \global\let\svgscale\undefined%
  \makeatother%
  \begin{picture}(1,0.41076452)%
    \put(0,0){\includegraphics[width=\unitlength]{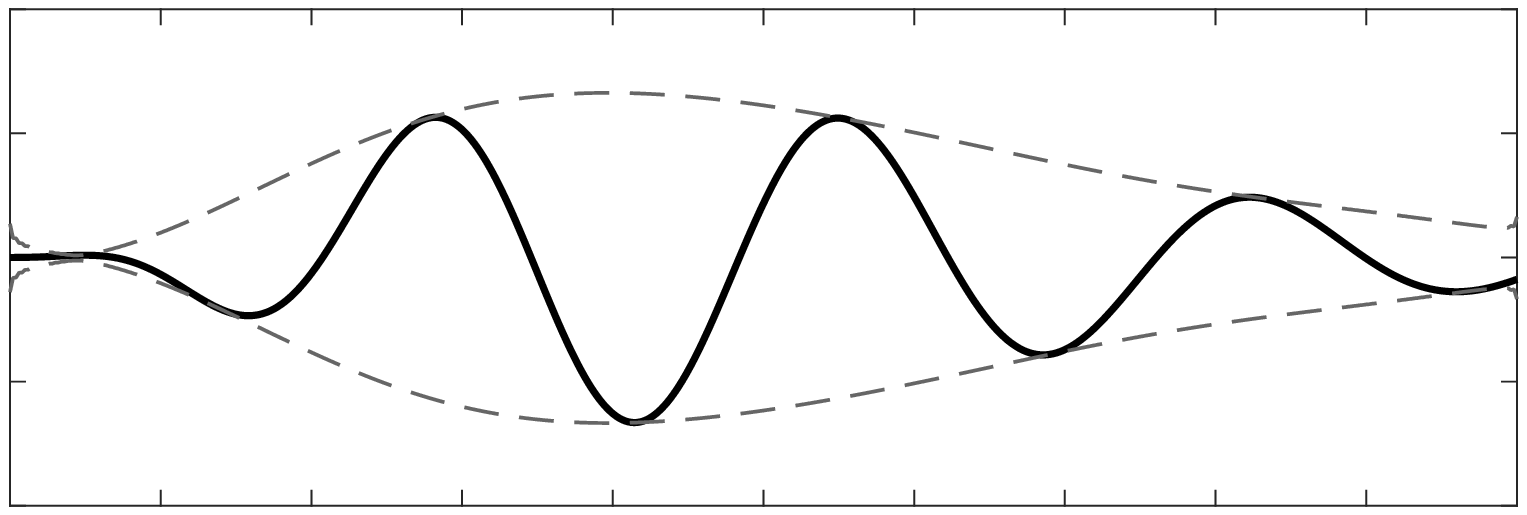}}%
      \begin{tiny}
    \put(0.085,0.035){\makebox(0,0)[lb]{\smash{0}}}
    \put(0.27,0.035){\makebox(0,0)[lb]{\smash{6}}}
    \put(0.43,0.035){\makebox(0,0)[lb]{\smash{12}}}
    \put(0.60,0.035){\makebox(0,0)[lb]{\smash{18}}}
    \put(0.77,0.035){\makebox(0,0)[lb]{\smash{24}}}
    \put(0.95,0.035){\makebox(0,0)[lb]{\smash{30}}}
    \put(0.015,0.06){\makebox(0,0)[lb]{\smash{-0.01}}} 
    \put(0.067,0.2){\makebox(0,0)[lb]{\smash{0}}}
    \put(0.02,0.34){\makebox(0,0)[lb]{\smash{0.01}}}
    \end{tiny}
    \put(0,0.2){\rotatebox{90}{\makebox(0,0)[lb]{\smash{h(t)}}}}%
    \put(0.1,0.37){\makebox(0,0)[lb]{\smash{Imp. resp. ($\omega_i=115.7$ Hz, $B_i=37.2$ Hz)}}}%

  \end{picture}%
\endgroup%

%% file: figures/method/gammatone/impulse-high-paper.eps_tex
\begingroup%
  \makeatletter%
  \providecommand\color[2][]{%
    \errmessage{(Inkscape) Color is used for the text in Inkscape, but the package 'color.sty' is not loaded}%
    \renewcommand\color[2][]{}%
  }%
  \providecommand\transparent[1]{%
    \errmessage{(Inkscape) Transparency is used (non-zero) for the text in Inkscape, but the package 'transparent.sty' is not loaded}%
    \renewcommand\transparent[1]{}%
  }%
  \providecommand\rotatebox[2]{#2}%
  \ifx\svgwidth\undefined%
    \setlength{\unitlength}{180bp}%
    \ifx\svgscale\undefined%
      \relax%
    \else%
      \setlength{\unitlength}{\unitlength * \real{\svgscale}}%
    \fi%
  \else%
    \setlength{\unitlength}{\svgwidth}%
  \fi%
  \global\let\svgwidth\undefined%
  \global\let\svgscale\undefined%
  \makeatother%
  \begin{picture}(1,0.41024523)%
    \put(0,0){\includegraphics[width=\unitlength]{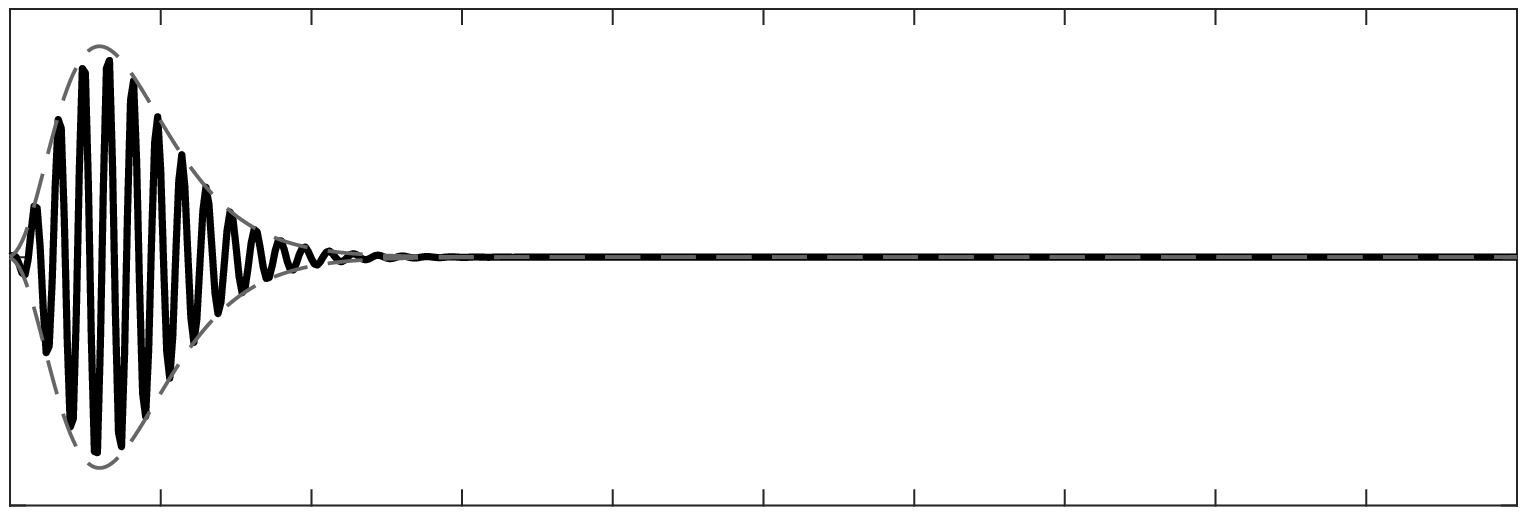}}%
    \begin{tiny}
    \put(0.085,0.035){\makebox(0,0)[lb]{\smash{0}}}
    \put(0.27,0.035){\makebox(0,0)[lb]{\smash{6}}}
    \put(0.43,0.035){\makebox(0,0)[lb]{\smash{12}}}
    \put(0.60,0.035){\makebox(0,0)[lb]{\smash{18}}}
    \put(0.77,0.035){\makebox(0,0)[lb]{\smash{24}}}
    \put(0.95,0.035){\makebox(0,0)[lb]{\smash{30}}}
    \put(0.015,0.06){\makebox(0,0)[lb]{\smash{-0.05}}} 
    \put(0.067,0.2){\makebox(0,0)[lb]{\smash{0}}}
    \put(0.02,0.34){\makebox(0,0)[lb]{\smash{0.05}}}
    \end{tiny}
   \put(0.48,-0.01){\makebox(0,0)[lb]{\smash{t (ms)}}}%
    \put(0,0.2){\rotatebox{90}{\makebox(0,0)[lb]{\smash{h(t)}}}}%
    \put(0.1,0.37){\makebox(0,0)[lb]{\smash{Imp. resp. ($\omega_i=1.96$ KHz, $B_i=0.24$ KHz)}}}%
    
  \end{picture}%
\endgroup%

%% file: figures/method/gammatone/frequency-resp-paper.eps_tex
\begingroup%
  \makeatletter%
  \providecommand\color[2][]{%
    \errmessage{(Inkscape) Color is used for the text in Inkscape, but the package 'color.sty' is not loaded}%
    \renewcommand\color[2][]{}%
  }%
  \providecommand\transparent[1]{%
    \errmessage{(Inkscape) Transparency is used (non-zero) for the text in Inkscape, but the package 'transparent.sty' is not loaded}%
    \renewcommand\transparent[1]{}%
  }%
  \providecommand\rotatebox[2]{#2}%
  \ifx\svgwidth\undefined%
    \setlength{\unitlength}{180bp}%
    \ifx\svgscale\undefined%
      \relax%
    \else%
      \setlength{\unitlength}{\unitlength * \real{\svgscale}}%
    \fi%
  \else%
    \setlength{\unitlength}{\svgwidth}%
  \fi%
  \global\let\svgwidth\undefined%
  \global\let\svgscale\undefined%
  \makeatother%
  \begin{picture}(1,0.81698018)%
    \put(0,0){\includegraphics[width=\unitlength]{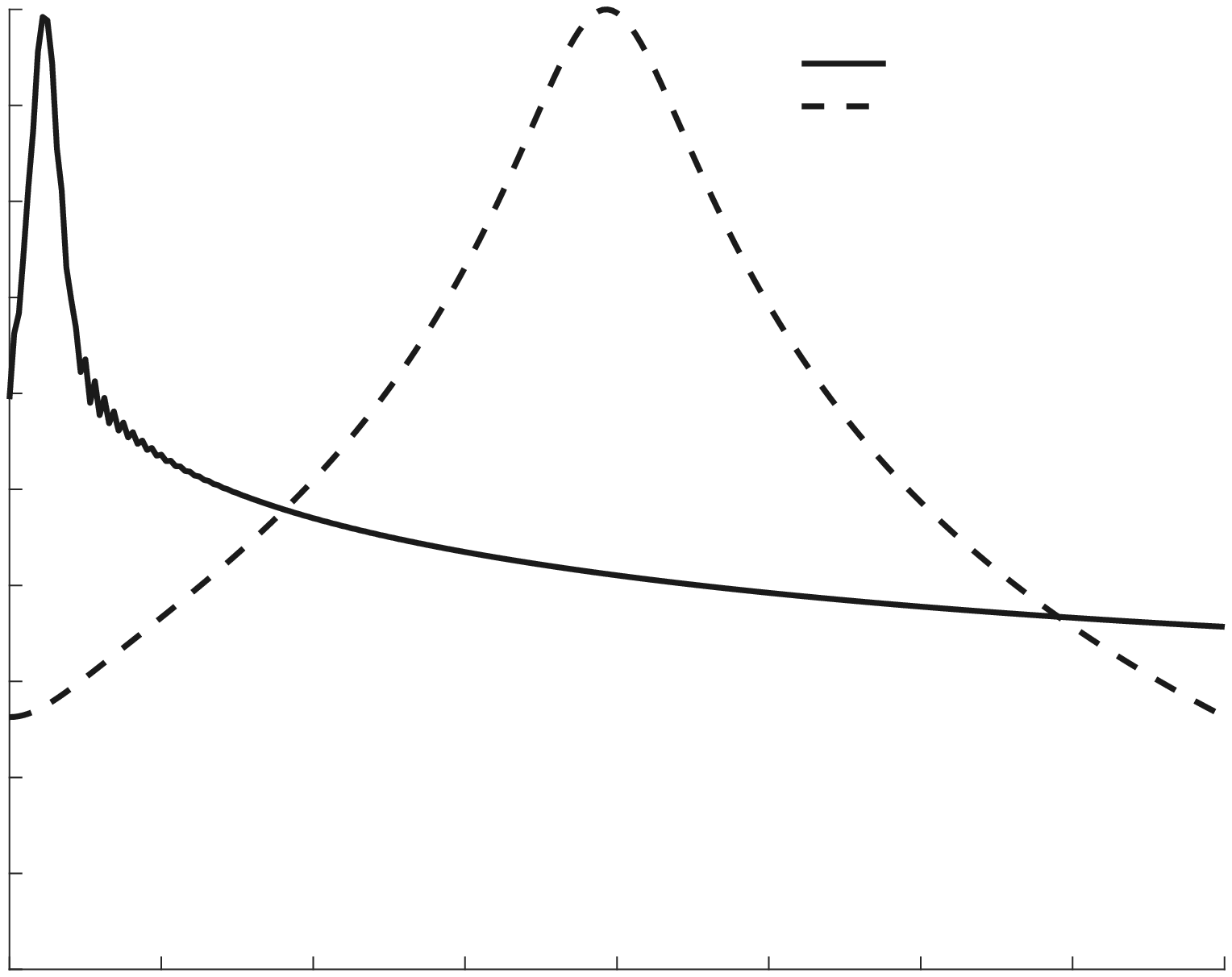}}%

    \put(0.46,-0.005){\makebox(0,0)[lb]{\smash{$\omega$ (KHz)}}}%
    \put(0,0.38){\color[rgb]{0.14901961,0.14901961,0.14901961}\rotatebox{90}{\makebox(0,0)[lb]{\smash{H(w) ($dB$)}}}}%
    \put(0.38,0.79){\makebox(0,0)[lb]{\smash{Frequency response}}}%
    
    \begin{tiny}
    \put(0.08,0.04){\makebox(0,0)[lb]{\smash{$0$}}} 
    \put(0.18,0.04){\makebox(0,0)[lb]{\smash{$0.5$}}}
    \put(0.31,0.04){\makebox(0,0)[lb]{\smash{$1$}}}    
    \put(0.41,0.04){\makebox(0,0)[lb]{\smash{$1.5$}}}
    \put(0.53,0.04){\makebox(0,0)[lb]{\smash{$2$}}}
    \put(0.63,0.04){\makebox(0,0)[lb]{\smash{$2.5$}}}
    \put(0.75,0.04){\makebox(0,0)[lb]{\smash{$3$}}}
    \put(0.85,0.04){\makebox(0,0)[lb]{\smash{$3.5$}}}
    \put(0.97,0.04){\makebox(0,0)[lb]{\smash{$4$}}}
    
    \put(0.02,0.065){\makebox(0,0)[lb]{\smash{-100}}}
    \put(0.03,0.13){\makebox(0,0)[lb]{\smash{-90}}}
    \put(0.03,0.20){\makebox(0,0)[lb]{\smash{-80}}}
    \put(0.03,0.27){\makebox(0,0)[lb]{\smash{-70}}}
    \put(0.03,0.34){\makebox(0,0)[lb]{\smash{-60}}}
    \put(0.03,0.41){\makebox(0,0)[lb]{\smash{-50}}}
    \put(0.03,0.48){\makebox(0,0)[lb]{\smash{-40}}}
    \put(0.03,0.55){\makebox(0,0)[lb]{\smash{-30}}}
    \put(0.03,0.62){\makebox(0,0)[lb]{\smash{-20}}}
    \put(0.03,0.69){\makebox(0,0)[lb]{\smash{-10}}}
    \put(0.049,0.76){\makebox(0,0)[lb]{\smash{0}}}

    \put(0.74,0.727){\color[rgb]{0,0,0}\makebox(0,0)[lb]{\smash{$\omega_i=115.7$ Hz}}}%
    \put(0.74,0.695){\color[rgb]{0,0,0}\makebox(0,0)[lb]{\smash{$\omega_i=1.96$ KHz}}}%
    \end{tiny}
    
  \end{picture}%
\endgroup%

%% file: figures/method/prototype-paper.eps_tex
\begingroup%
  \makeatletter%
  \providecommand\color[2][]{%
    \errmessage{(Inkscape) Color is used for the text in Inkscape, but the package 'color.sty' is not loaded}%
    \renewcommand\color[2][]{}%
  }%
  \providecommand\transparent[1]{%
    \errmessage{(Inkscape) Transparency is used (non-zero) for the text in Inkscape, but the package 'transparent.sty' is not loaded}%
    \renewcommand\transparent[1]{}%
  }%
  \providecommand\rotatebox[2]{#2}%
  \ifx\svgwidth\undefined%
    \setlength{\unitlength}{210bp}%
    \ifx\svgscale\undefined%
      \relax%
    \else%
      \setlength{\unitlength}{\unitlength * \real{\svgscale}}%
    \fi%
  \else%
    \setlength{\unitlength}{\svgwidth}%
  \fi%
  \global\let\svgwidth\undefined%
  \global\let\svgscale\undefined%
  \makeatother%
  \begin{picture}(1,0.5620915)%
    \put(0,0){\includegraphics[width=\unitlength]{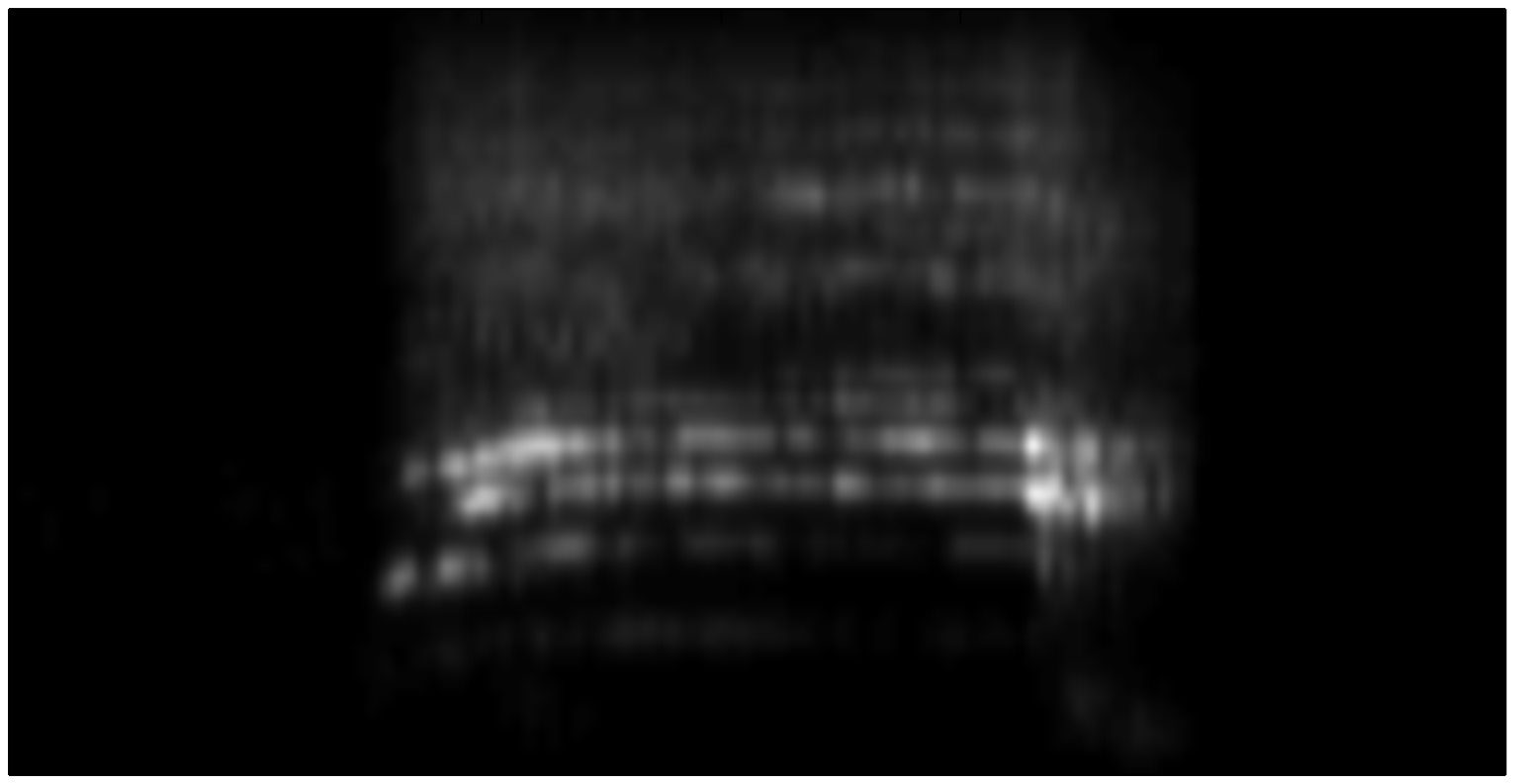}}%
    \put(0.02529248,0.24){\color[rgb]{0,0,0}\rotatebox{90}{\makebox(0,0)[lb]{\smash{freq. (KHz)}}}}%
    \put(0.44,-0.01){\color[rgb]{0,0,0}\makebox(0,0)[lb]{\smash{time (seconds)}}}%
    \put(0.32,0.53){\color[rgb]{0,0,0}\makebox(0,0)[lb]{\smash{Prototype sound (scream)}}}%
    \put(0.1,0.030){\color[rgb]{0,0,0}\makebox(0,0)[lb]{\smash{0}}}%
    \put(0.22085989,0.028){\color[rgb]{0,0,0}\makebox(0,0)[lb]{\smash{0.32}}}%
    \put(0.36332312,0.028){\color[rgb]{0,0,0}\makebox(0,0)[lb]{\smash{0.64}}}%
    \put(0.50578636,0.028){\color[rgb]{0,0,0}\makebox(0,0)[lb]{\smash{0.96}}}%
    \put(0.64808007,0.028){\color[rgb]{0,0,0}\makebox(0,0)[lb]{\smash{1.28}}}%
    \put(0.79616013,0.028){\color[rgb]{0,0,0}\makebox(0,0)[lb]{\smash{1.6}}}%
    \put(0.93283701,0.028){\color[rgb]{0,0,0}\makebox(0,0)[lb]{\smash{1.92}}}%
    \put(0.04,0.13038662){\color[rgb]{0,0,0}\makebox(0,0)[lb]{\smash{0.2}}}%
    \put(0.04,0.19705201){\color[rgb]{0,0,0}\makebox(0,0)[lb]{\smash{0.7}}}%
    \put(0.04,0.26387893){\color[rgb]{0,0,0}\makebox(0,0)[lb]{\smash{1.5}}}%
    \put(0.058,0.3378942){\color[rgb]{0,0,0}\makebox(0,0)[lb]{\smash{3}}}%
    \put(0.058,0.39753278){\color[rgb]{0,0,0}\makebox(0,0)[lb]{\smash{6}}}%
    \put(0.050,0.4643597){\color[rgb]{0,0,0}\makebox(0,0)[lb]{\smash{12}}}%
  \end{picture}%
\endgroup%

%% file: figures/method/allpeaks-paper.eps_tex
\begingroup%
  \makeatletter%
  \providecommand\color[2][]{%
    \errmessage{(Inkscape) Color is used for the text in Inkscape, but the package 'color.sty' is not loaded}%
    \renewcommand\color[2][]{}%
  }%
  \providecommand\transparent[1]{%
    \errmessage{(Inkscape) Transparency is used (non-zero) for the text in Inkscape, but the package 'transparent.sty' is not loaded}%
    \renewcommand\transparent[1]{}%
  }%
  \providecommand\rotatebox[2]{#2}%
  \ifx\svgwidth\undefined%
    \setlength{\unitlength}{210bp}%
    \ifx\svgscale\undefined%
      \relax%
    \else%
      \setlength{\unitlength}{\unitlength * \real{\svgscale}}%
    \fi%
  \else%
    \setlength{\unitlength}{\svgwidth}%
  \fi%
  \global\let\svgwidth\undefined%
  \global\let\svgscale\undefined%
  \makeatother%
  \begin{picture}(1,0.5620915)%
    \put(0,0){\includegraphics[width=\unitlength]{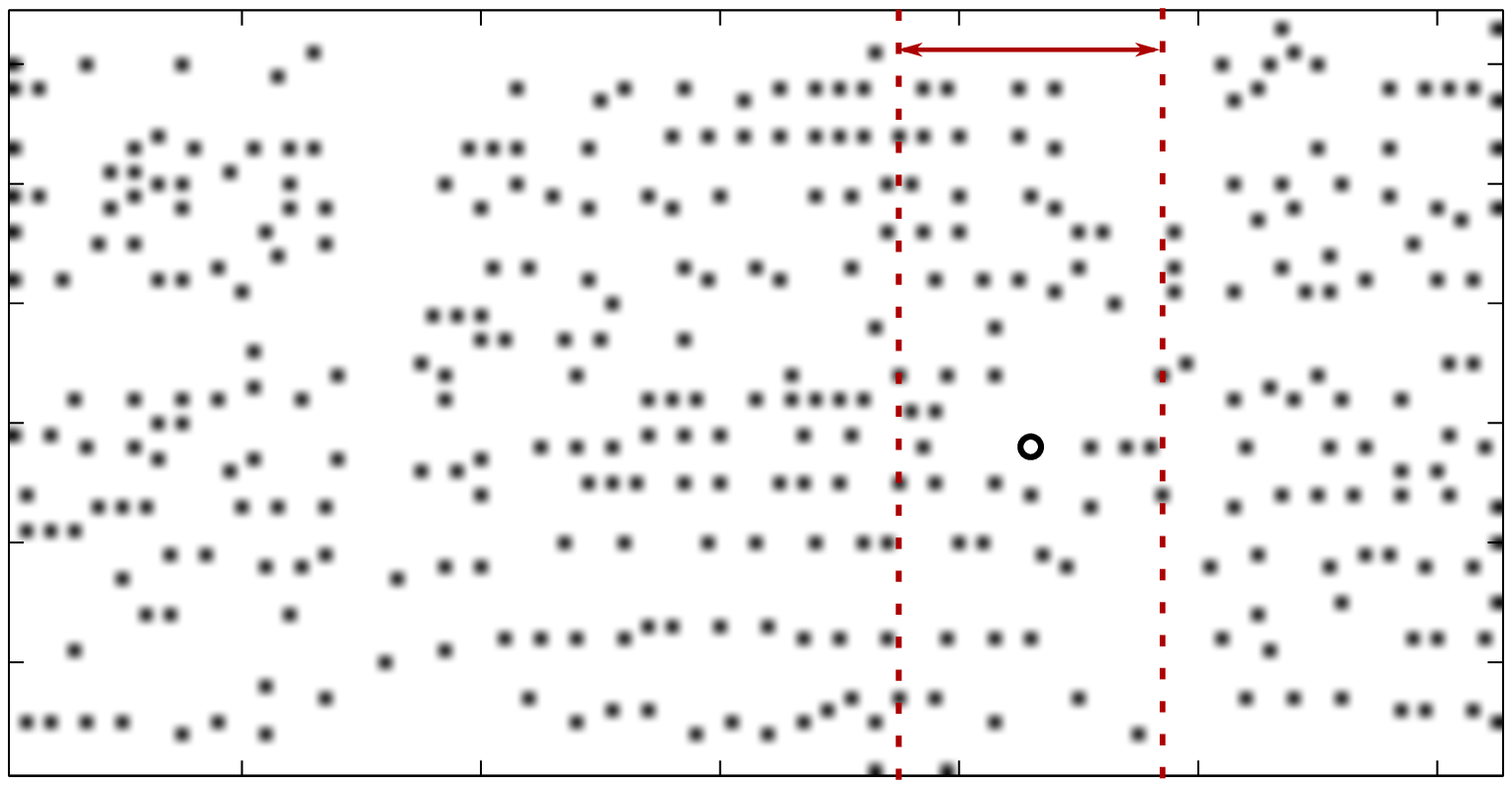}}%
       \put(0.02529248,0.24){\color[rgb]{0,0,0}\rotatebox{90}{\makebox(0,0)[lb]{\smash{freq. (KHz)}}}}%
    \put(0.44,-0.01){\color[rgb]{0,0,0}\makebox(0,0)[lb]{\smash{time (seconds)}}}%
    \put(0.15,0.53){\color[rgb]{0,0,0}\makebox(0,0)[lb]{\smash{Energy peaks and feature extractor support}}}%
    \put(0.1,0.033){\color[rgb]{0,0,0}\makebox(0,0)[lb]{\smash{0}}}%
    \put(0.22085989,0.03){\color[rgb]{0,0,0}\makebox(0,0)[lb]{\smash{0.32}}}%
    \put(0.36332312,0.03){\color[rgb]{0,0,0}\makebox(0,0)[lb]{\smash{0.64}}}%
    \put(0.50578636,0.03){\color[rgb]{0,0,0}\makebox(0,0)[lb]{\smash{0.96}}}%
    \put(0.64808007,0.03){\color[rgb]{0,0,0}\makebox(0,0)[lb]{\smash{1.28}}}%
    \put(0.79616013,0.03){\color[rgb]{0,0,0}\makebox(0,0)[lb]{\smash{1.6}}}%
    \put(0.93283701,0.03){\color[rgb]{0,0,0}\makebox(0,0)[lb]{\smash{1.92}}}%
    \put(0.040,0.13038662){\color[rgb]{0,0,0}\makebox(0,0)[lb]{\smash{0.2}}}%
    \put(0.040,0.19705201){\color[rgb]{0,0,0}\makebox(0,0)[lb]{\smash{0.7}}}%
    \put(0.040,0.26387893){\color[rgb]{0,0,0}\makebox(0,0)[lb]{\smash{1.5}}}%
    \put(0.059,0.3378942){\color[rgb]{0,0,0}\makebox(0,0)[lb]{\smash{3}}}%
    \put(0.059,0.39753278){\color[rgb]{0,0,0}\makebox(0,0)[lb]{\smash{6}}}%
    \put(0.050,0.4643597){\color[rgb]{0,0,0}\makebox(0,0)[lb]{\smash{12}}}%
  \end{picture}%
\endgroup%

%% file: figures/method/configuredpeaks-paper.eps_tex
\begingroup%
  \makeatletter%
  \providecommand\color[2][]{%
    \errmessage{(Inkscape) Color is used for the text in Inkscape, but the package 'color.sty' is not loaded}%
    \renewcommand\color[2][]{}%
  }%
  \providecommand\transparent[1]{%
    \errmessage{(Inkscape) Transparency is used (non-zero) for the text in Inkscape, but the package 'transparent.sty' is not loaded}%
    \renewcommand\transparent[1]{}%
  }%
  \providecommand\rotatebox[2]{#2}%
  \ifx\svgwidth\undefined%
    \setlength{\unitlength}{210bp}%
    \ifx\svgscale\undefined%
      \relax%
    \else%
      \setlength{\unitlength}{\unitlength * \real{\svgscale}}%
    \fi%
  \else%
    \setlength{\unitlength}{\svgwidth}%
  \fi%
  \global\let\svgwidth\undefined%
  \global\let\svgscale\undefined%
  \makeatother%
  \begin{picture}(1,0.5620915)%
    \put(0,0){\includegraphics[width=\unitlength]{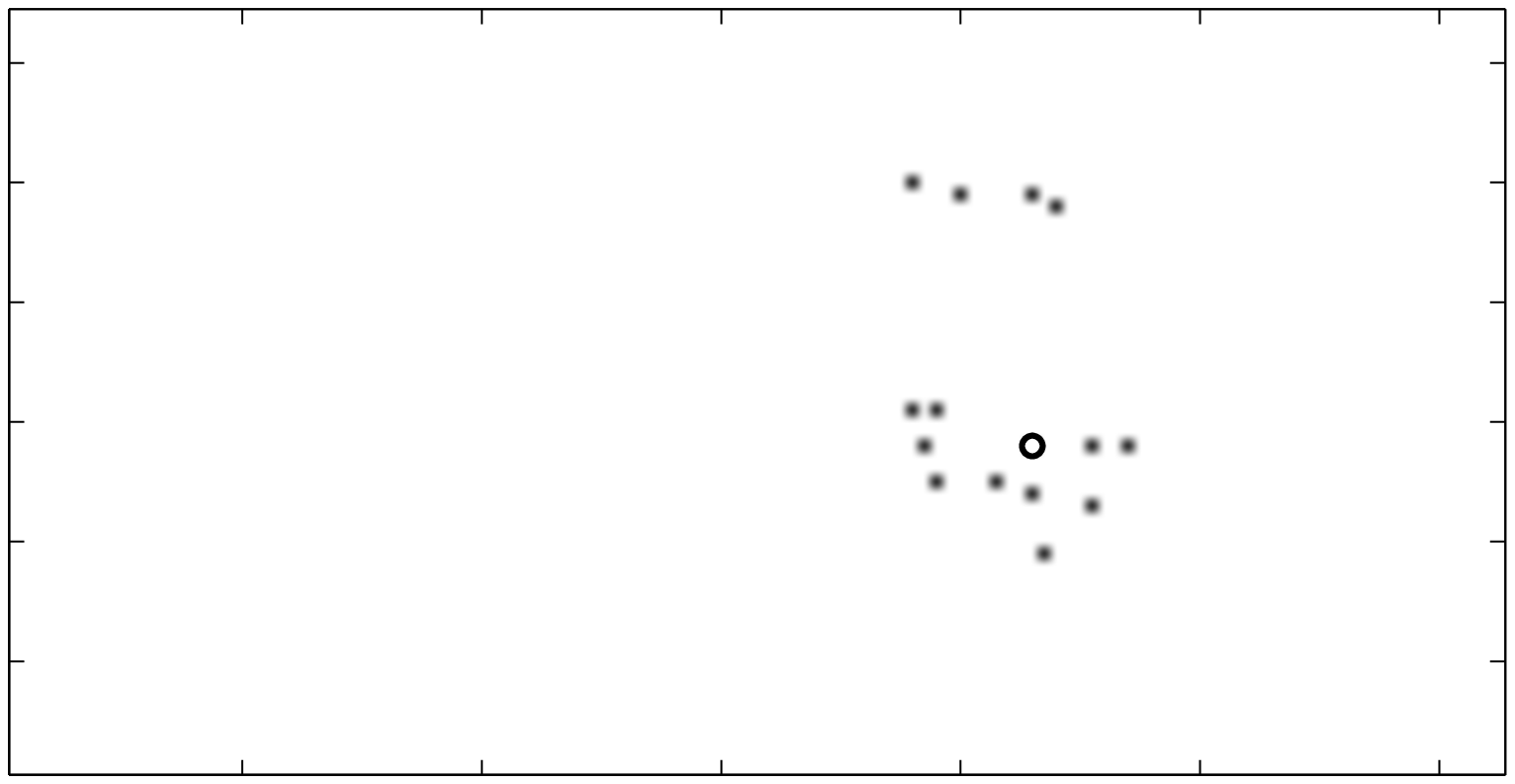}}%
    \put(0.02529248,0.24){\color[rgb]{0,0,0}\rotatebox{90}{\makebox(0,0)[lb]{\smash{freq. (KHz)}}}}%
    \put(0.44,-0.01){\color[rgb]{0,0,0}\makebox(0,0)[lb]{\smash{time (seconds)}}}%
    \put(0.24,0.53){\color[rgb]{0,0,0}\makebox(0,0)[lb]{\smash{Configured COPE feature extractor}}}%
    \put(0.1,0.033){\color[rgb]{0,0,0}\makebox(0,0)[lb]{\smash{0}}}%
    \put(0.22085989,0.03){\color[rgb]{0,0,0}\makebox(0,0)[lb]{\smash{0.32}}}%
    \put(0.36332312,0.03){\color[rgb]{0,0,0}\makebox(0,0)[lb]{\smash{0.64}}}%
    \put(0.50578636,0.03){\color[rgb]{0,0,0}\makebox(0,0)[lb]{\smash{0.96}}}%
    \put(0.64808007,0.03){\color[rgb]{0,0,0}\makebox(0,0)[lb]{\smash{1.28}}}%
    \put(0.79616013,0.03){\color[rgb]{0,0,0}\makebox(0,0)[lb]{\smash{1.6}}}%
    \put(0.93283701,0.03){\color[rgb]{0,0,0}\makebox(0,0)[lb]{\smash{1.92}}}%
    \put(0.040,0.13038662){\color[rgb]{0,0,0}\makebox(0,0)[lb]{\smash{0.2}}}%
    \put(0.040,0.19705201){\color[rgb]{0,0,0}\makebox(0,0)[lb]{\smash{0.7}}}%
    \put(0.040,0.26387893){\color[rgb]{0,0,0}\makebox(0,0)[lb]{\smash{1.5}}}%
    \put(0.058,0.3378942){\color[rgb]{0,0,0}\makebox(0,0)[lb]{\smash{3}}}%
    \put(0.058,0.39753278){\color[rgb]{0,0,0}\makebox(0,0)[lb]{\smash{6}}}%
    \put(0.051,0.4643597){\color[rgb]{0,0,0}\makebox(0,0)[lb]{\smash{12}}}%
  \end{picture}%
\endgroup%

%% file: figures/comparison/events-det.eps_tex
\begingroup%
  \makeatletter%
  \providecommand\color[2][]{%
    \errmessage{(Inkscape) Color is used for the text in Inkscape, but the package 'color.sty' is not loaded}%
    \renewcommand\color[2][]{}%
  }%
  \providecommand\transparent[1]{%
    \errmessage{(Inkscape) Transparency is used (non-zero) for the text in Inkscape, but the package 'transparent.sty' is not loaded}%
    \renewcommand\transparent[1]{}%
  }%
  \providecommand\rotatebox[2]{#2}%
  \ifx\svgwidth\undefined%
    \setlength{\unitlength}{175bp}%
    \ifx\svgscale\undefined%
      \relax%
    \else%
      \setlength{\unitlength}{\unitlength * \real{\svgscale}}%
    \fi%
  \else%
    \setlength{\unitlength}{\svgwidth}%
  \fi%
  \global\let\svgwidth\undefined%
  \global\let\svgscale\undefined%
  \makeatother%
  \begin{picture}(1,0.9906324)%
    \put(0,0){\includegraphics[width=\unitlength]{events-det.eps}}%
    \put(0.14,0.05){\makebox(0,0)[lb]{\smash{0.1 }}}%
    \put(0.218,0.05){\makebox(0,0)[lb]{\smash{0.2 }}}%
    \put(0.29071617,0.05){\makebox(0,0)[lb]{\smash{0.5 }}}%
    \put(0.36306847,0.05){\makebox(0,0)[lb]{\smash{1 }}}%
    \put(0.43134579,0.05){\makebox(0,0)[lb]{\smash{2 }}}%
    \put(0.53365967,0.05){\makebox(0,0)[lb]{\smash{5 }}}%
    \put(0.62115854,0.05){\makebox(0,0)[lb]{\smash{10 }}}%
    \put(0.73128118,0.05){\makebox(0,0)[lb]{\smash{20 }}}%
    \put(0.87844612,0.05){\makebox(0,0)[lb]{\smash{40 }}}%
    \put(0.07,0.133){\makebox(0,0)[lb]{\smash{0.1 }}}%
    \put(0.07,0.186131){\makebox(0,0)[lb]{\smash{0.2 }}}%
    \put(0.07,0.26181631){\makebox(0,0)[lb]{\smash{0.5 }}}%
    \put(0.09156415,0.32408451){\makebox(0,0)[lb]{\smash{1 }}}%
    \put(0.09156415,0.39236183){\makebox(0,0)[lb]{\smash{2 }}}%
    \put(0.09156415,0.49467571){\makebox(0,0)[lb]{\smash{5 }}}%
    \put(0.082,0.5855792){\makebox(0,0)[lb]{\smash{10 }}}%
    \put(0.082,0.69570183){\makebox(0,0)[lb]{\smash{20 }}}%
    \put(0.082,0.84286678){\makebox(0,0)[lb]{\smash{40 }}}%
    \put(0.27,0.01){\makebox(0,0)[lb]{\smash{False Alarm probability ($\%$)}}}%
    \put(0.05517969,0.33){\rotatebox{90}{\makebox(0,0)[lb]{\smash{Miss probability ($\%$)}}}}%
    \put(0.35,0.93416912){\makebox(0,0)[lb]{\smash{MIVIA audio events }}}%
    \put(0.85,0.86){\makebox(0,0)[lb]{\smash{COPE}}}%
    \put(0.85,0.81){\makebox(0,0)[lb]{\smash{$bof_h$}}}%
    \put(0.85,0.76){\makebox(0,0)[lb]{\smash{$bof_s$}}}%
  \end{picture}%
\endgroup%

%% file: figures/comparison/roads-det.eps_tex
\begingroup%
  \makeatletter%
  \providecommand\color[2][]{%
    \errmessage{(Inkscape) Color is used for the text in Inkscape, but the package 'color.sty' is not loaded}%
    \renewcommand\color[2][]{}%
  }%
  \providecommand\transparent[1]{%
    \errmessage{(Inkscape) Transparency is used (non-zero) for the text in Inkscape, but the package 'transparent.sty' is not loaded}%
    \renewcommand\transparent[1]{}%
  }%
  \providecommand\rotatebox[2]{#2}%
  \ifx\svgwidth\undefined%
    \setlength{\unitlength}{175bp}%
    \ifx\svgscale\undefined%
      \relax%
    \else%
      \setlength{\unitlength}{\unitlength * \real{\svgscale}}%
    \fi%
  \else%
    \setlength{\unitlength}{\svgwidth}%
  \fi%
  \global\let\svgwidth\undefined%
  \global\let\svgscale\undefined%
  \makeatother%
  \begin{picture}(1,0.9906324)%
    \put(0,0){\includegraphics[width=\unitlength]{roads-det.eps}}%
    \put(0.14,0.05){\makebox(0,0)[lb]{\smash{0.1 }}}%
    \put(0.218,0.05){\makebox(0,0)[lb]{\smash{0.2 }}}%
    \put(0.29071617,0.05){\makebox(0,0)[lb]{\smash{0.5 }}}%
    \put(0.36306847,0.05){\makebox(0,0)[lb]{\smash{1 }}}%
    \put(0.43134579,0.05){\makebox(0,0)[lb]{\smash{2 }}}%
    \put(0.53365967,0.05){\makebox(0,0)[lb]{\smash{5 }}}%
    \put(0.62115854,0.05){\makebox(0,0)[lb]{\smash{10 }}}%
    \put(0.73128118,0.05){\makebox(0,0)[lb]{\smash{20 }}}%
    \put(0.87844612,0.05){\makebox(0,0)[lb]{\smash{40 }}}%
    \put(0.07,0.133){\makebox(0,0)[lb]{\smash{0.1 }}}%
    \put(0.07,0.186131){\makebox(0,0)[lb]{\smash{0.2 }}}%
    \put(0.07,0.26181631){\makebox(0,0)[lb]{\smash{0.5 }}}%
    \put(0.09156415,0.32408451){\makebox(0,0)[lb]{\smash{1 }}}%
    \put(0.09156415,0.39236183){\makebox(0,0)[lb]{\smash{2 }}}%
    \put(0.09156415,0.49467571){\makebox(0,0)[lb]{\smash{5 }}}%
    \put(0.082,0.5855792){\makebox(0,0)[lb]{\smash{10 }}}%
    \put(0.082,0.69570183){\makebox(0,0)[lb]{\smash{20 }}}%
    \put(0.082,0.84286678){\makebox(0,0)[lb]{\smash{40 }}}%
    \put(0.27,0.01){\makebox(0,0)[lb]{\smash{False Alarm probability ($\%$)}}}%
    \put(0.05517969,0.33){\rotatebox{90}{\makebox(0,0)[lb]{\smash{Miss probability ($\%$)}}}}%
    \put(0.35,0.93416912){\makebox(0,0)[lb]{\smash{MIVIA road events }}}%
    \put(0.28158961,0.2876333){\makebox(0,0)[lb]{\smash{COPE}}}%
    \put(0.28158961,0.23877694){\makebox(0,0)[lb]{\smash{BARK}}}%
    \put(0.28158961,0.1901224){\makebox(0,0)[lb]{\smash{MFCC}}}%
    \put(0.28158961,0.14126845){\makebox(0,0)[lb]{\smash{\cite{Carletti13}}}}%
  \end{picture}%
\endgroup%

%% file: figures/results/rocversions-paper.eps_tex
\begingroup%
  \makeatletter%
  \providecommand\color[2][]{%
    \errmessage{(Inkscape) Color is used for the text in Inkscape, but the package 'color.sty' is not loaded}%
    \renewcommand\color[2][]{}%
  }%
  \providecommand\transparent[1]{%
    \errmessage{(Inkscape) Transparency is used (non-zero) for the text in Inkscape, but the package 'transparent.sty' is not loaded}%
    \renewcommand\transparent[1]{}%
  }%
  \providecommand\rotatebox[2]{#2}%
  \ifx\svgwidth\undefined%
    \setlength{\unitlength}{220bp}%
    \ifx\svgscale\undefined%
      \relax%
    \else%
      \setlength{\unitlength}{\unitlength * \real{\svgscale}}%
    \fi%
  \else%
    \setlength{\unitlength}{\svgwidth}%
  \fi%
  \global\let\svgwidth\undefined%
  \global\let\svgscale\undefined%
  \makeatother%
  \footnotesize
  \begin{picture}(1,0.80961399)%
    \put(0,0){\includegraphics[width=\unitlength]{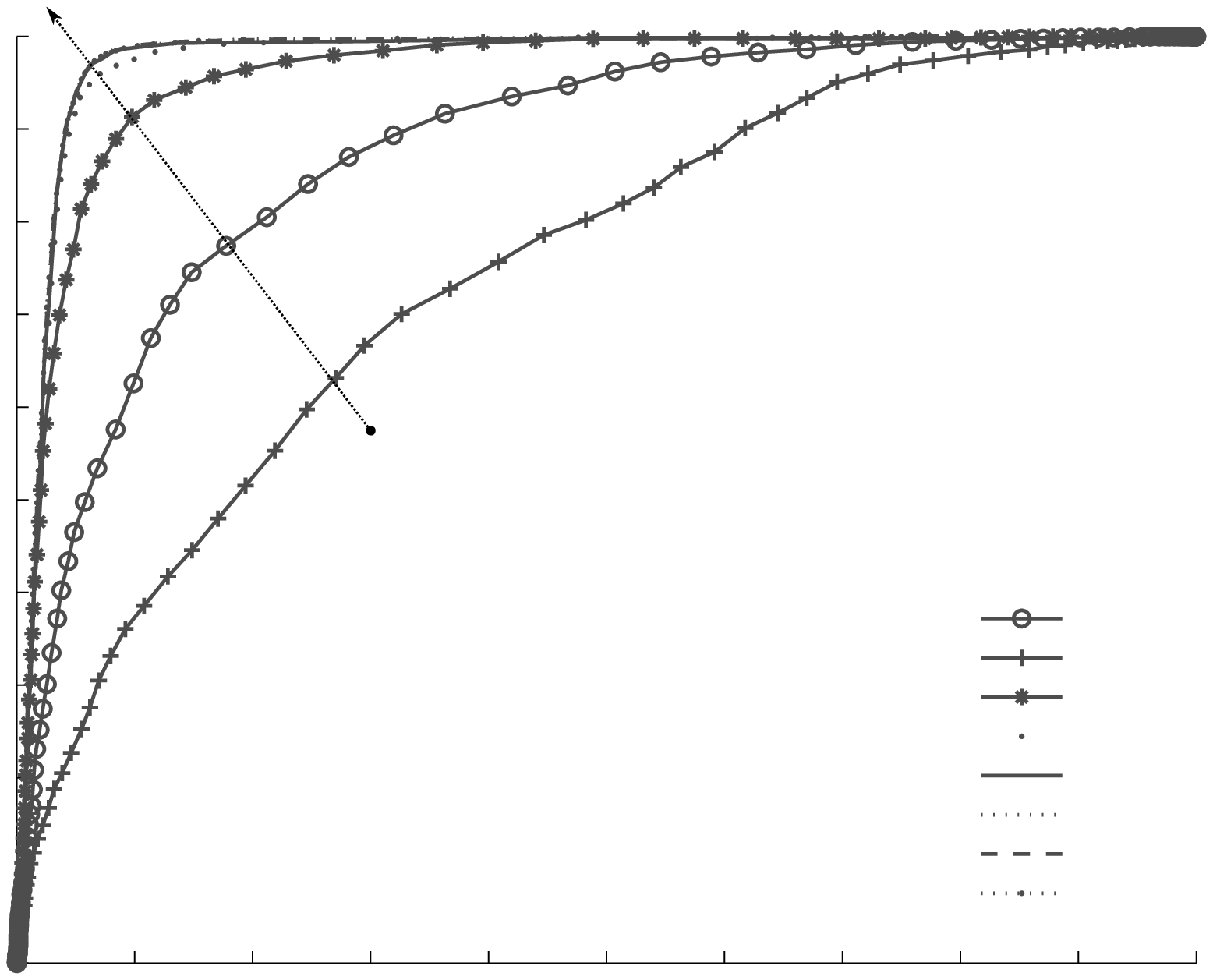}}%
    \put(0.0742051,0.02){\color[rgb]{0,0,0}\makebox(0,0)[lb]{\smash{0}}}%
    \put(0.15590173,0.02){\color[rgb]{0,0,0}\makebox(0,0)[lb]{\smash{0.1}}}%
    \put(0.24642398,0.02){\color[rgb]{0,0,0}\makebox(0,0)[lb]{\smash{0.2}}}%
    \put(0.33694624,0.02){\color[rgb]{0,0,0}\makebox(0,0)[lb]{\smash{0.3}}}%
    \put(0.42747057,0.02){\color[rgb]{0,0,0}\makebox(0,0)[lb]{\smash{0.4}}}%
    \put(0.51799282,0.02){\color[rgb]{0,0,0}\makebox(0,0)[lb]{\smash{0.5}}}%
    \put(0.60851508,0.02){\color[rgb]{0,0,0}\makebox(0,0)[lb]{\smash{0.6}}}%
    \put(0.69903941,0.02){\color[rgb]{0,0,0}\makebox(0,0)[lb]{\smash{0.7}}}%
    \put(0.78956166,0.02){\color[rgb]{0,0,0}\makebox(0,0)[lb]{\smash{0.8}}}%
    \put(0.88008392,0.02){\color[rgb]{0,0,0}\makebox(0,0)[lb]{\smash{0.9}}}%
    \put(0.97926106,0.02){\color[rgb]{0,0,0}\makebox(0,0)[lb]{\smash{1}}}%
    \put(0.04,0.05429436){\color[rgb]{0,0,0}\makebox(0,0)[lb]{\smash{0}}}%
    \put(0.02,0.1254319){\color[rgb]{0,0,0}\makebox(0,0)[lb]{\smash{0.1}}}%
    \put(0.02,0.19656943){\color[rgb]{0,0,0}\makebox(0,0)[lb]{\smash{0.2}}}%
    \put(0.02,0.26770697){\color[rgb]{0,0,0}\makebox(0,0)[lb]{\smash{0.3}}}%
    \put(0.02,0.3388445){\color[rgb]{0,0,0}\makebox(0,0)[lb]{\smash{0.4}}}%
    \put(0.02,0.40980965){\color[rgb]{0,0,0}\makebox(0,0)[lb]{\smash{0.5}}}%
    \put(0.02,0.48094718){\color[rgb]{0,0,0}\makebox(0,0)[lb]{\smash{0.6}}}%
    \put(0.02,0.55208472){\color[rgb]{0,0,0}\makebox(0,0)[lb]{\smash{0.7}}}%
    \put(0.02,0.62322225){\color[rgb]{0,0,0}\makebox(0,0)[lb]{\smash{0.8}}}%
    \put(0.02,0.69435979){\color[rgb]{0,0,0}\makebox(0,0)[lb]{\smash{0.9}}}%
    \put(0.04,0.76532285){\color[rgb]{0,0,0}\makebox(0,0)[lb]{\smash{1}}}%
    \put(0.51141494,-0.015){\color[rgb]{0,0,0}\makebox(0,0)[lb]{\smash{FPR}}}%
    \put(0,0.39648148){\color[rgb]{0,0,0}\rotatebox{90}{\makebox(0,0)[lb]{\smash{TPR}}}}%
    \put(0.88856642,0.31894053){\color[rgb]{0,0,0}\makebox(0,0)[lb]{\smash{$0dB$}}}%
    \put(0.88856642,0.28882391){\color[rgb]{0,0,0}\makebox(0,0)[lb]{\smash{$-5dB$}}}%
    \put(0.88856642,0.25870729){\color[rgb]{0,0,0}\makebox(0,0)[lb]{\smash{$5dB$}}}%
    \put(0.88856642,0.22859067){\color[rgb]{0,0,0}\makebox(0,0)[lb]{\smash{$10dB$}}}%
    \put(0.88856642,0.19847405){\color[rgb]{0,0,0}\makebox(0,0)[lb]{\smash{$15dB$}}}%
    \put(0.88856642,0.16835701){\color[rgb]{0,0,0}\makebox(0,0)[lb]{\smash{$20dB$}}}%
    \put(0.88856642,0.13824039){\color[rgb]{0,0,0}\makebox(0,0)[lb]{\smash{$25dB$}}}%
    \put(0.88856642,0.10812377){\color[rgb]{0,0,0}\makebox(0,0)[lb]{\smash{$30dB$}}}%
    \put(0.12023152,0.78542553){\color[rgb]{0,0,0}\makebox(0,0)[lb]{\smash{SNR}}}%
  \end{picture}%
\endgroup%

%% file: figures/discussion/sound-paper.eps_tex
\begingroup%
  \makeatletter%
  \providecommand\color[2][]{%
    \errmessage{(Inkscape) Color is used for the text in Inkscape, but the package 'color.sty' is not loaded}%
    \renewcommand\color[2][]{}%
  }%
  \providecommand\transparent[1]{%
    \errmessage{(Inkscape) Transparency is used (non-zero) for the text in Inkscape, but the package 'transparent.sty' is not loaded}%
    \renewcommand\transparent[1]{}%
  }%
  \providecommand\rotatebox[2]{#2}%
  \ifx\svgwidth\undefined%
    \setlength{\unitlength}{180bp}%
    \ifx\svgscale\undefined%
      \relax%
    \else%
      \setlength{\unitlength}{\unitlength * \real{\svgscale}}%
    \fi%
  \else%
    \setlength{\unitlength}{\svgwidth}%
  \fi%
  \global\let\svgwidth\undefined%
  \global\let\svgscale\undefined%
  \makeatother%
  \begin{picture}(1,0.57639846)%
    \put(0,0){\includegraphics[width=\unitlength]{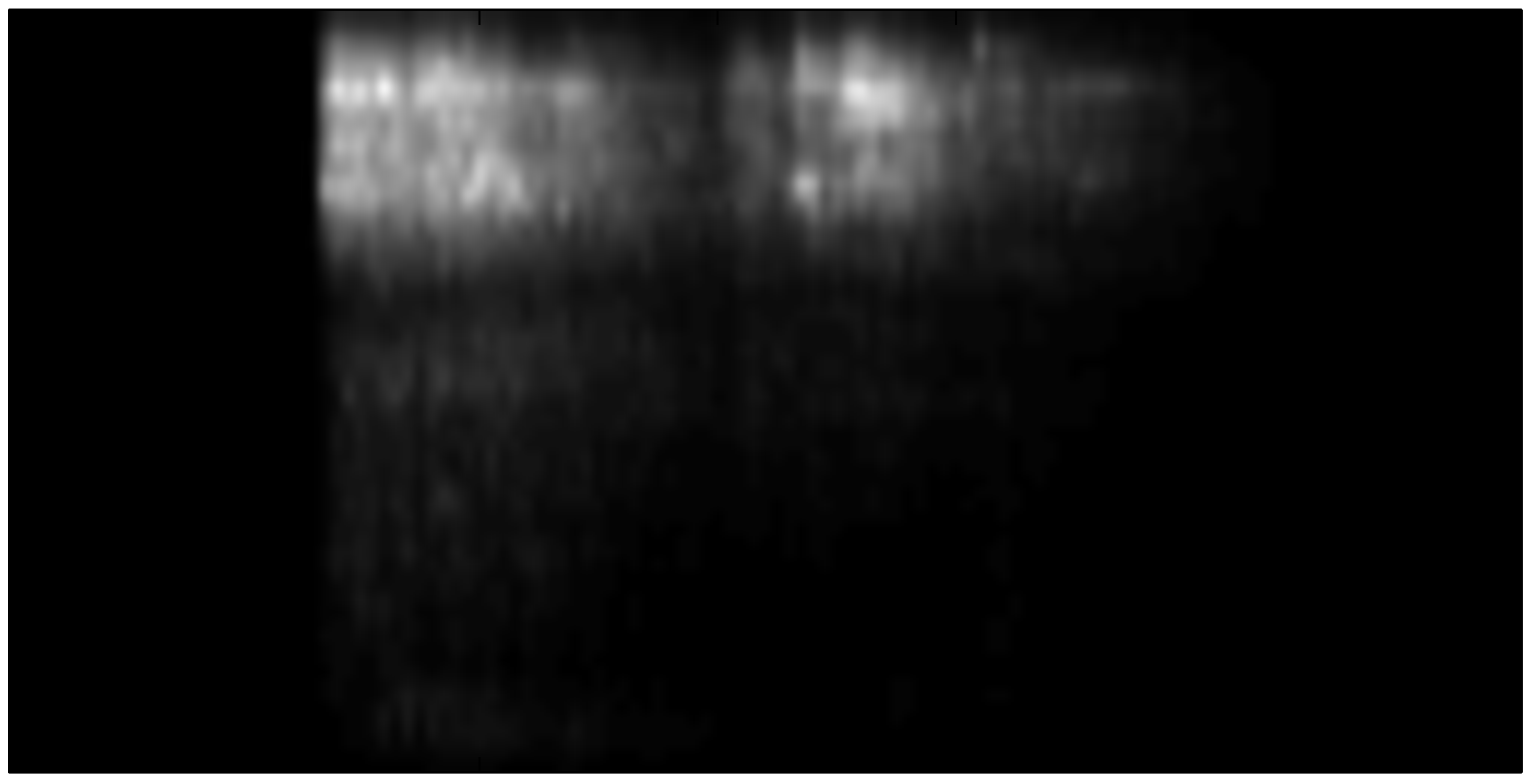}}%
    \put(0.4,0.001){\makebox(0,0)[lb]{\smash{time (seconds)}}}%
    \put(0.05,0.13038662){\makebox(0,0)[lb]{\smash{0.2}}}%
    \put(0.05,0.19705201){\makebox(0,0)[lb]{\smash{0.7}}}%
    \put(0.05,0.26387893){\makebox(0,0)[lb]{\smash{1.5}}}%
    \put(0.079,0.33070585){\makebox(0,0)[lb]{\smash{3}}}%
    \put(0.079,0.39753278){\makebox(0,0)[lb]{\smash{6}}}%
    \put(0.06,0.4643597){\makebox(0,0)[lb]{\smash{12}}}%
    \put(0.22,0.53){\makebox(0,0)[lb]{\smash{Glass breaking at 30\textit{dB} SNR}}}%
    \put(0.03,0.22){\rotatebox{90}{\makebox(0,0)[lb]{\smash{freq. (KHz)}}}}%
    \put(0.1,0.03){\makebox(0,0)[lb]{\smash{0}}}%
    \put(0.22237115,0.038){\makebox(0,0)[lb]{\smash{0.32}}}%
    \put(0.356025,0.038){\makebox(0,0)[lb]{\smash{0.64}}}%
    \put(0.48951923,0.038){\makebox(0,0)[lb]{\smash{0.96}}}%
    \put(0.62317308,0.038){\makebox(0,0)[lb]{\smash{1.28}}}%
    \put(0.76211538,0.038){\makebox(0,0)[lb]{\smash{1.6}}}%
    \put(0.89032115,0.038){\makebox(0,0)[lb]{\smash{1.92}}}%
  \end{picture}%
\endgroup%

%% file: figures/discussion/resp-paper.eps_tex
\begingroup%
  \makeatletter%
  \providecommand\color[2][]{%
    \errmessage{(Inkscape) Color is used for the text in Inkscape, but the package 'color.sty' is not loaded}%
    \renewcommand\color[2][]{}%
  }%
  \providecommand\transparent[1]{%
    \errmessage{(Inkscape) Transparency is used (non-zero) for the text in Inkscape, but the package 'transparent.sty' is not loaded}%
    \renewcommand\transparent[1]{}%
  }%
  \providecommand\rotatebox[2]{#2}%
  \ifx\svgwidth\undefined%
    \setlength{\unitlength}{180bp}%
    \ifx\svgscale\undefined%
      \relax%
    \else%
      \setlength{\unitlength}{\unitlength * \real{\svgscale}}%
    \fi%
  \else%
    \setlength{\unitlength}{\svgwidth}%
  \fi%
  \global\let\svgwidth\undefined%
  \global\let\svgscale\undefined%
  \makeatother%
  \begin{picture}(1,0.56038012)%
    \put(0.045,0.02){\includegraphics[width=\unitlength]{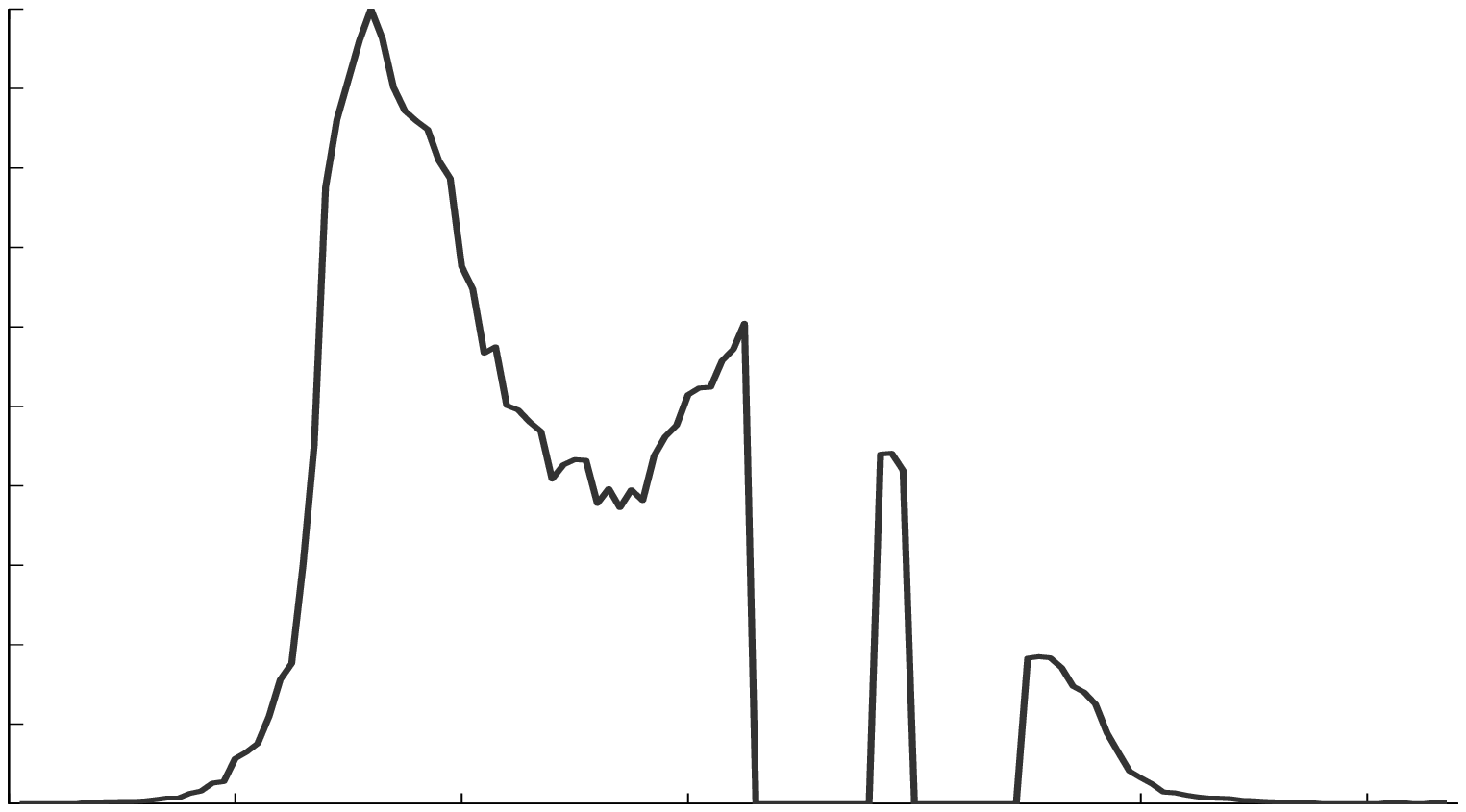}}%
   \put(0.1,0.028){\makebox(0,0)[lb]{\smash{0}}}%
    \put(0.21,0.028){\makebox(0,0)[lb]{\smash{0.32}}}%
    \put(0.35,0.028){\makebox(0,0)[lb]{\smash{0.64}}}%
    \put(0.48,0.028){\makebox(0,0)[lb]{\smash{0.96}}}%
    \put(0.62,0.028){\makebox(0,0)[lb]{\smash{1.28}}}%
    \put(0.76,0.028){\makebox(0,0)[lb]{\smash{1.6}}}%
    \put(0.90,0.028){\makebox(0,0)[lb]{\smash{1.92}}}%
    \put(0.45,0.0){\makebox(0,0)[lb]{\smash{time (seconds)}}}%

    \put(0.01,0.31){\rotatebox{90}{\makebox(0,0)[lb]{\smash{$r(t)$}}}}%

    \put(0.07,0.06){\makebox(0,0)[lb]{\smash{0}}}%
    \put(0.03,0.155){\makebox(0,0)[lb]{\smash{0.2}}}%
    \put(0.03,0.25){\makebox(0,0)[lb]{\smash{0.4}}}%
    \put(0.03,0.345){\makebox(0,0)[lb]{\smash{0.6}}}%
    \put(0.03,0.438){\makebox(0,0)[lb]{\smash{0.8}}}%
    \put(0.07,0.53){\makebox(0,0)[lb]{\smash{1}}}%
    \put(0.1,0.56){\makebox(0,0)[lb]{\smash{Response of the COPE feature extractor}}}%
  \end{picture}%
\endgroup%

%% file: figures/discussion/respdetail-paper.eps_tex
\begingroup%
  \makeatletter%
  \providecommand\color[2][]{%
    \errmessage{(Inkscape) Color is used for the text in Inkscape, but the package 'color.sty' is not loaded}%
    \renewcommand\color[2][]{}%
  }%
  \providecommand\transparent[1]{%
    \errmessage{(Inkscape) Transparency is used (non-zero) for the text in Inkscape, but the package 'transparent.sty' is not loaded}%
    \renewcommand\transparent[1]{}%
  }%
  \providecommand\rotatebox[2]{#2}%
  \ifx\svgwidth\undefined%
    \setlength{\unitlength}{180bp}%
    \ifx\svgscale\undefined%
      \relax%
    \else%
      \setlength{\unitlength}{\unitlength * \real{\svgscale}}%
    \fi%
  \else%
    \setlength{\unitlength}{\svgwidth}%
  \fi%
  \global\let\svgwidth\undefined%
  \global\let\svgscale\undefined%
  \makeatother%
  \begin{picture}(1,0.56066982)%
    \put(0.0375,0.02){\includegraphics[width=\unitlength]{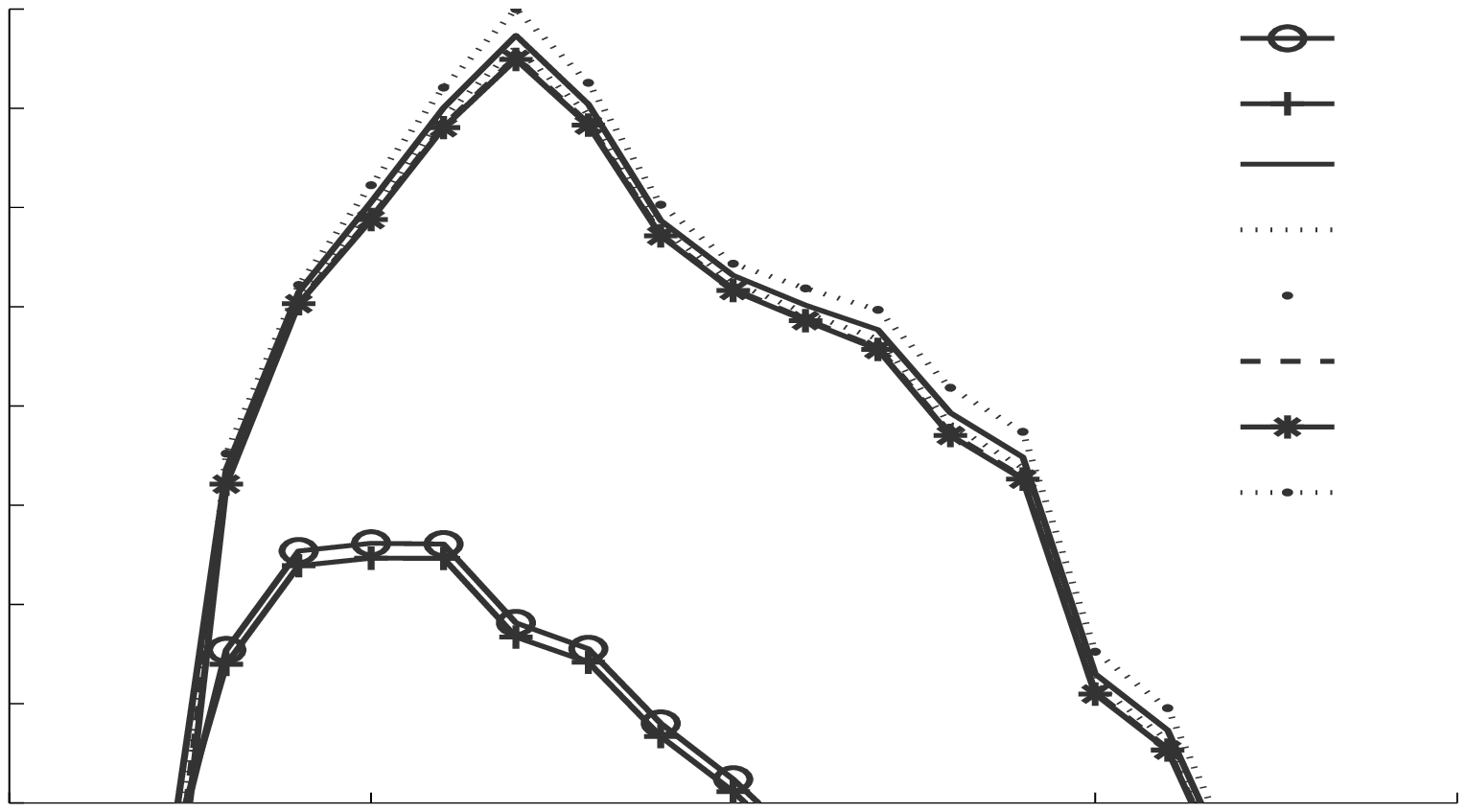}}%
    \put(0.1,0.035){\makebox(0,0)[lb]{\smash{0.4}}}%
    \put(0.285,0.035){\makebox(0,0)[lb]{\smash{0.445}}}%
    \put(0.505,0.035){\makebox(0,0)[lb]{\smash{0.49}}}%
    \put(0.715,0.035){\makebox(0,0)[lb]{\smash{0.535}}}%
    \put(0.935,0.035){\makebox(0,0)[lb]{\smash{0.58}}}%
	\put(0.45,0.0){\makebox(0,0)[lb]{\smash{time (seconds)}}}%
	
    \put(0.043,0.064){\makebox(0,0)[lb]{\smash{0.6}}}%
    \put(0.043,0.175){\makebox(0,0)[lb]{\smash{0.7}}}%
    \put(0.043,0.295){\makebox(0,0)[lb]{\smash{0.8}}}%
    \put(0.043,0.412){\makebox(0,0)[lb]{\smash{0.9}}}%
    \put(0.073,0.53){\makebox(0,0)[lb]{\smash{1}}}%
    
        \put(0.01,0.31){\rotatebox{90}{\makebox(0,0)[lb]{\smash{$r(t)$}}}}%

    \put(0.28,0.564){\makebox(0,0)[lb]{\smash{Response at various SNRs}}}%
    \put(0.93,0.52){\makebox(0,0)[lb]{\smash{0dB}}}%
    \put(0.935,0.48){\makebox(0,0)[lb]{\smash{-5dB}}}%
    \put(0.93,0.445){\makebox(0,0)[lb]{\smash{5dB}}}%
    \put(0.93,0.404){\makebox(0,0)[lb]{\smash{10dB}}}%
    \put(0.93,0.363){\makebox(0,0)[lb]{\smash{15dB}}}%
    \put(0.93,0.324){\makebox(0,0)[lb]{\smash{20dB}}}%
    \put(0.93,0.283){\makebox(0,0)[lb]{\smash{25dB}}}%
    \put(0.93,0.242){\makebox(0,0)[lb]{\smash{30dB}}}%
  \end{picture}%
\endgroup%